\documentclass[11pt]{report}
\usepackage{amsmath,amsfonts,amssymb,epsfig,color,psfrag,young}
\usepackage[vcentermath]{youngtab}
\usepackage{rotating}
\begin{document}
\title{On P vs. NP, Geometric Complexity Theory, and the Riemann 
Hypothesis}
\author{
Dedicated to Sri Ramakrishna \\ \\
Ketan D. Mulmuley 
 \\
The University of Chicago
\\  \\
http://ramakrishnadas.cs.uchicago.edu \\ \\
Technical Report, Computer Science Department,\\
The University of Chicago 
}

\maketitle

\newtheorem{prop}{Proposition}[section]
\newtheorem{claim}[prop]{Claim}
\newtheorem{goal}[prop]{Goal}
\newtheorem{theorem}[prop]{Theorem}
\newtheorem{metathesis}[prop]{Metathesis}
\newtheorem{mainpoint}{Main Point}
\newtheorem{hypo}[prop]{Hypothesis}
\newtheorem{guess}[prop]{Guess}
\newtheorem{problem}[prop]{Problem}
\newtheorem{axiom}[prop]{Axiom}
\newtheorem{question}[prop]{Question}
\newtheorem{remark}[prop]{Remark}
\newtheorem{lemma}[prop]{Lemma}
\newtheorem{claimedlemma}[prop]{Claimed Lemma}
\newtheorem{claimedtheorem}[prop]{Claimed Theorem}
\newtheorem{cor}[prop]{Corollary}
\newtheorem{defn}[prop]{Definition}
\newtheorem{ex}[prop]{Example}
\newtheorem{conj}[prop]{Conjecture}
\newtheorem{obs}[prop]{Observation}
\newtheorem{phyp}[prop]{Positivity Hypothesis}
\newcommand{\bitlength}[1]{\langle #1 \rangle}
\newcommand{\ca}[1]{{\cal #1}}
\newcommand{\floor}[1]{{\lfloor #1 \rfloor}}
\newcommand{\ceil}[1]{{\lceil #1 \rceil}}
\newcommand{\gt}[1]{{\langle  #1 |}}
\newcommand{\C}{\mathbb{C}}
\newcommand{\N}{\mathbb{N}}
\newcommand{\R}{\mathbb{R}}
\newcommand{\Z}{\mathbb{Z}}
\newcommand{\frcgc}[5]{\left(\begin{array}{ll} #1 &  \\ #2 & | #4 \\ #3 & | #5
\end{array}\right)}

\newcommand{\cgc}[6]{\left(\begin{array}{ll} #1 ;& \quad #3\\ #2 ; & \quad #4
\end{array}\right| \left. \begin{array}{l} #5 \\ #6 \end{array} \right)}

\newcommand{\wigner}[6]
{\left(\begin{array}{ll} #1 ;& \quad #3\\ #2 ; & \quad #4
\end{array}\right| \left. \begin{array}{l} #5 \\ #6 \end{array} \right)}

\newcommand{\rcgc}[9]{\left(\begin{array}{ll} #1 & \quad #4\\ #2  & \quad #5
\\ #3 &\quad #6
\end{array}\right| \left. \begin{array}{l} #7 \\ #8 \\#9 \end{array} \right)}

\newcommand{\srcgc}[4]{\left(\begin{array}{ll} #1 & \\ #2 & | #4  \\ #3 & |
\end{array}\right)}

\newcommand{\arr}[2]{\left(\begin{array}{l} #1 \\ #2   \end{array} \right)}
\newcommand{\unshuffle}[1]{\langle #1 \rangle}
\newcommand{\ignore}[1]{}
\newcommand{\f}[2]{{\frac {#1} {#2}}}
\newcommand{\tableau}[5]{
\begin{array}{ccc} 
#1 & #2  &#3 \\
#4 & #5 
\end{array}}
\newcommand{\embed}[1]{{#1}^\phi}
\newcommand{\stab}{{\mbox {stab}}}
\newcommand{\perm}{{\mbox {perm}}}
\newcommand{\trace}{{\mbox {trace}}}
\newcommand{\polylog}{{\mbox {polylog}}}
\newcommand{\sign}{{\mbox {sign}}}
\newcommand{\proj}{{\mbox {Proj}}}
\newcommand{\poly}{{\mbox {poly}}}
\newcommand{\std}{{\mbox {std}}}
\newcommand{\m}{\mbox}
\newcommand{\formula}{{\mbox {Formula}}}
\newcommand{\circuit}{{\mbox {Circuit}}}
\newcommand{\sgn}{{\mbox {sgn}}}
\newcommand{\core}{{\mbox {core}}}
\newcommand{\orbit}{{\mbox {orbit}}}
\newcommand{\cycle}{{\mbox {cycle}}}
\newcommand{\ideal}{{\mbox {ideal}}}
\newcommand{\qed}{{\mbox {Q.E.D.}}}
\newcommand{\proof}{\noindent {\em Proof: }}
\newcommand{\weight}{{\mbox {wt}}}
\newcommand{\tab}{{\mbox {Tab}}}
\newcommand{\level}{{\mbox {level}}}
\newcommand{\vol}{{\mbox {vol}}}
\newcommand{\vect}{{\mbox {Vect}}}
\newcommand{\val}{{\mbox {wt}}}
\newcommand{\sym}{{\mbox {Sym}}}
\newcommand{\convex}{{\mbox {convex}}}
\newcommand{\spec}{{\mbox {spec}}}
\newcommand{\strong}{{\mbox {strong}}}
\newcommand{\adm}{{\mbox {Adm}}}
\newcommand{\eval}{{\mbox {eval}}}
\newcommand{\for}{{\quad \mbox {for}\ }}
\newcommand{\Q}{Q}
\newcommand{\mand}{{\quad \mbox {and}\ }}
\newcommand{\invlim}{{\mbox {lim}_\leftarrow}}
\newcommand{\directlim}{{\mbox {lim}_\rightarrow}}
\newcommand{\sformal}{{\cal S}^{\mbox f}}
\newcommand{\vformal}{{\cal V}^{\mbox f}}
\newcommand{\crystal}{\mbox{crystal}}
\newcommand{\conje}{\mbox{\bf Conj}}
\newcommand{\graph}{\mbox{graph}}
\newcommand{\ind}{\mbox{index}}

\newcommand{\rank}{\mbox{rank}}
\newcommand{\id}{\mbox{id}}
\newcommand{\str}{\mbox{string}}
\newcommand{\RSK}{\mbox{RSK}}
\newcommand{\wt}{\mbox{wt}}
\setlength{\unitlength}{.75in}

\subsection*{Abstract}
Geometric complexity
theory (GCT) is an approach to the $P$ vs. $NP$ and related problems
suggested in a series of  articles we call GCTlocal \cite{GCTpram},
GCT1-8 \cite{GCT1}-\cite{GCT8}, and GCTflip
\cite{GCTflip}.
A  high level overview of this research plan and the results 
obtained so far  was presented  in 
a series of three lectures in the Institute of Advanced study, Princeton,
Feb 9-11, 2009.  This article contains the material covered in
those lectures after some revision, and gives a mathematical overview of GCT.
No background in algebraic geometry, representation theory
or quantum groups is assumed. 
For those who are interested in a short mathematical overview, the first 
lecture (chapter) of this article gives this. 
The video lectures for this series   are available at:

http://video.ias.edu/csdm/pvsnp

They may be a helpful supplement to  this article.

\section*{Introduction}

\begin{figure} 
\begin{center}
\epsfig{file=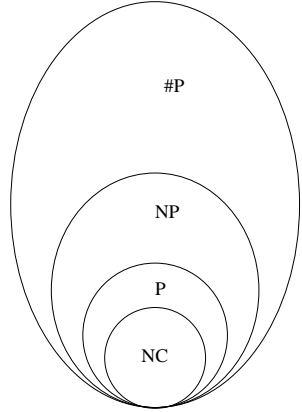, scale=.6}
\end{center}
      \caption{Complexity Classes}
      \label{fig:complexityclasses}
\end{figure}

This article   gives a mathematical  overview of
geometric complexity theory (GCT), an
approach  towards the fundamental lower bound problems  in complexity theory, 
such as (Figure~\ref{fig:complexityclasses}):

\noindent (1) 
The $P$ vs. $NP$ problem \cite{cook,karp,levin}: 
 show that $P \not = NP$;

\noindent (2) 
The  $\#P$ vs. $NC$ problem \cite{valiant}: show that 
$\#P \not = NC$.

\noindent (3) The $P$ vs. $NC$ problem: show that $P \not = NC$.

We  focus here on only the nonuniform versions of the above problems 
in characteristic zero; i.e., when
the underlying field of computation is of characteristic zero, 
say $\Q$ or $\C$--what this means will be explained below.
The additional problems that
need to be addressed when the underlying field of computation is finite
would be discussed in  GCT11.

The nonuniform
characteristic zero version of the $P\not = NC$ conjecture (in fact,
something stronger) was already proved  in GCTlocal. We shall refer
to it as the $P\not = NC$ result without bit operations.
It says that the max flow problem cannot be solved in the 
PRAM model without bit operations in $\polylog(N)$ time using 
$\poly(N)$ processors where $N$ is the bitlength of the input.
This  may be considered to be the first unconditional lower bound result of
GCT, because, though it can be stated in purely elementary combinatorial
terms, being a formal implication of the $P\not = NC$ conjecture, its
proof is intrinsically geometric, and no elementary proof is known so far. 
Furthermore, its proof  technique may be considered to be a weaker (local)
form of the {\em flip}, the basic guiding strategy of GCT, which  was 
refined and formalized much later in GCTflip. 
This  was the begining of  this  geometric approach in complexity theory.
The later work in GCT-- the subject of this overview--focusses on the other 
two problems above, namely
the $P$ vs. $NP$ and $\#P$ vs. $NC$ problems.

The nonuniform (characteristic zero) version of the $\#P$ vs. $NC$ problem
is also known as the permanent vs. determinant problem \cite{valiant}. 
It  is to show that $\perm(X)$, the permanent of an $n\times n$ variable 
matrix $X$, cannot be represented 
linearly  as $\det(Y)$, the determinant of 
an $m\times m$ matrix $Y$,
if $m=\poly(n)$, or more generally, $m=2^{\log^a n}$, 
for a fixed constant $a>0$, 
and $n\rightarrow \infty$. 
By linear representation, we mean
the  entries   of $Y$ are  (possibly nonhomogeneous) 
linear functions of the entries of $X$. 
There is an analogous characteristic zero version of
the $P$ vs. $NP$ problem defined in GCT1, where the role of
the permanent is played by an appropriate (co)-NP complete function and
the role of the determinant is played by an appropriate $P$-complete function.
The main results of GCT for the $\#P$ vs. $NC$ problem in 
characteristic zero also extend to the $P$ vs. $NP$ problem in characteristic
zero. But here we 
concentrate  on only the former problem,
since this illustrates all the basic ideas.

The complementary article \cite{GCTexplicit} gives a complexity-theoretic
overview of GCT. It describes   the main complexity theoretic
barrier towards these problems 
called the {\em complexity barrier} and the defining
strategy of GCT for crossing it called the {\em flip} [GCT6,GCTflip]: 
which is to go 
for {\em explicit proofs}. By an explicit proof 
we mean a proof that provides proof certificates of hardness  for the 
hard function under consideration that are short (of polynomial size) 
and easy to verify (in polynomial time). This barrier turns out to be
extremely formidable and is the root cause of all difficulties 
in these problems. Nonelementary techniques are brought into GCT precisely
to cross this barrier. It is not discussed in these lectures. 
The goal here is to describe the basic ideas of GCT at a concrete mathematical
level without getting into such meta issues. But the readers who wish
to know the need for the nonelementary techniques in GCT before getting
into any mathematics   may wish to read that
article before this one. On the other hand, the readers who would rather avoid 
meta issues before getting a concrete mathematical
picture  may wish to read this article first. We  leave the
 choice to the readers.

The original IAS lectures stated a lower bound called a 
{\em weak form  of the $\#P$ vs. $NC$
problem}. This is a special case of a more general result 
which we shall call {\em a mathematical form of the $\#P \not = NC$ 
conjecture} (Section~\ref{sweaklec1}). 
It follows easily from  basic results in geometric invariant theory.
The article \cite{boris} showed that the weak form  stated in the IAS 
lectures is  too weak because it has a direct elementary (linear algebraic)
proof.
Hence in this article it has been replaced with the 
mathematical form of the $\#P\not = NC$ conjecture mentioned above; cf. 
Section~\ref{sweaklec1}. 
We cannot prove this  mathematical form by elementary linear algebraic proof.

The rest of this article is organized in the form of three chapters,
one per  lecture. The first gives a short mathematical 
overview of the basic plan of GCT,  
which is elaborated in the next two lectures.

\noindent {\em Acknowledgement:} The author is grateful to 
Avi Wigderson for arranging the lectures and the hospitality,
to the authors of \cite{boris} for pointing out weakness of the 
lower bound stated during the lectures, and to Shrawan Kumar for
bringing the reference \cite{ariki} to his attention and 
helpful discussions.

\chapter{Basic plan} 
We now outline the basic  plan of GCT focussing on
the  permanent vs. determinant problem in characteristic zero. 

\section{Characterization by symmetries} \label{scharsym}
We begin by observing that 
the permanent and the determinant are 
{\em exceptional} polynomial functions. By exceptional we mean
that they are  completely characterized by the symmetries in the 
following sense.

Let  $Y$ be a variable $m\times m$ matrix.
Let $\sym^m(Y)$ be the space of homogeneous forms of degree $m$ 
in the $m^2$ variable entries of $Y$. Then, by the  classical
representation theory,  $\det(Y)$ is the only form in $\sym^m(Y)$ 
such that, for any $A,B \in GL_m(\C)$ with $\det(A)\det(B)=1$, 

\noindent  {\bf (D):} $\det(Y)=\det(A Y^* B)$,

where $Y^*=Y$ or $Y^t$. 
Thus $\det(Y)$ is completely characterized 
by its symmetries, and hence, is  exceptional. We shall refer 
to this characteristic property of the determinant as property (D)
henceforth.

Similarly, $\perm(X)$ is the only form 
in the space of forms of degree $n$ in the entries of $X$ such that, 
for any diagonal or permutation matrices $A,B$,

\noindent  {\bf (P):} $\perm(X)=\perm(AX^*B)$, 

where $X^*=X$ or $X^t$  with obvious constraints 
on the product of the diagonal entries of $A$ and $B$ when they are diagonal.
Thus $\perm(X)$ is also
completely characterized   by its symmetries, and hence, is
exceptional.
We shall refer 
to this characteristic property of the permanent as property (P)
henceforth.

A basic idea [GCT1]  is to a get a handle on 
the permanent vs. determinant problem 
by exploiting exceptional nature of these polynomials--i.e.,
their characteristic properties (P) and  (D).
Representation theory and 
algebraic geometry enter inevitably into  the study of these properties,
because
to understand symmetries representation theory (of groups of symmetries)
becomes indispensible, and to understand  deeper properties 
of representations  algebraic geometry becomes indispensible.

\section{A mathematical form of the  $\#P \not = NC$ conjecture} 
\label{sweaklec1}
To show how these characteristic properties can be exploited, 
we now state one application of 
GCT in the form of a concrete lower bound result--namely a 
mathematical form of the 
$\#P \not = NC$ conjecture (Theorem~\ref{tgenweaklec1} below)--before
going any further.

We begin by observing that 
the permanent vs. determinant conjecture clearly implies  that 
$\perm(X)$ of any $n \times n$ variable matrix $X$ can not be represented 
as an $NC$-computable polynomial in the traces of $\bar X^j$, $j \ge 0$,
$\bar X=B X C$
for any (possibly singular) $n\times n$ matrices $B$ and $C$, since
$X^j$ can be computed fast in parallel. This can be proved unconditionally.
In fact, something stronger.

\begin{prop}  \label{pweaklec1}
There do not exist (possibly singular) $n\times n$ complex
matrices $B$ and $C$  and a polynomial 
$e(w_0,\ldots, w_n)$ such that $\perm(X) = g(BXC)$, 
where $g(X)=e(\trace(X^0), \trace(X), \ldots, \trace(X^n))$. 
\end{prop}
This was referred to as 
the weak form of the $\#P$ vs. $NC$ problem in the original IAS lecture. 
The article \cite{boris} showed that this is too  weak 
by giving an  elementary linear algebraic proof  \cite{boris}.

We now state a  more general lower bound, which
was not stated in the IAS lecture, and which does not have such 
an elementary linear algebraic proof. For that we need a definition.

\begin{defn} \label{dgenperm}
A polynomial function $p(X_1,\ldots,X_k)$ (of any degree)
in the entries of $k$ $n\times n$ variable matrices $X_1,\ldots,X_k$  
is called a  {\em generalized permanent} if it has exactly 
the same symmetries as that of the permanent; i.e.,
for all nonsingular $n\times n$ matrices $U_i$ and $V_i$, $i \le k$, 
\[ p(U_1 X_1 V_1,\ldots, U_k X_k V_k)=p(X_1,\ldots,X_k) 
\mbox{\ iff\ } \perm(U_i X V_i)=\perm (X) \quad \forall i.\] 
\end{defn}

A precise description of the symmetries of the permanent  is given
by the property (P). Hence, $U_i$ and $V_i$ above have to be permutation
or diagonal matrices (with obvious constraints on the product of their 
diagonal entries), or 
product of such matrices.
When $k=1$ and $n$ is arbitrary,
there is just one generalized permanent of degree $n$, namely
the usual permanent itself. 
At the other extreme, 
when $n=1$ and $k$ is arbitrary, every function in $k$ variables 
is a generalized permanent. 
For general  $n$ and $k$, almost any polynomial 
in $\perm(X_i)$'s, $i \le k$, is  a generalized permanent, but 
there are many others besides these.
For general degrees, 
the dimension of the 
space spanned by  generalized permanents can be  exponential
in $n$; cf. Section~\ref{sweak}. 
In general,  the space of generalized permanents has a highly nontrivial
structure that  is intimately linked  to some
fundamental problems of representation theory;
cf. Section~\ref{sweak} and  [GCT6].

Now we have the following:

\begin{obs}[Implication of the nonuniform $\#P \not = NC$ conjecture]

Assuming the nonuniform $\#P \not = NC$ conjecture in characteristic zero,
no $\#P$-complete generalized permanent  $p(X_1,\ldots,X_k)$ of $\poly(n,k)$ 
degree can be expressed as an $NC$-computable polynomial function of the 
traces of  $\bar X_i^j$, $1 \le i \le k$, $j=\poly(n,k)$, where 
$\bar X_i=B_i X_i C_i$,  $i \le k$,  for any $n\times n$ complex (possibly
singular) matrices $B_i$ and $C_i$.
\end{obs}
(Here $X_i^j$ are  clearly $NC$-computable).

When $n=1$ and $k$ is arbitrary, this implication is  equivalent 
to the original nonuniform $\#P \not = NC$ conjecture (in characteristic 
zero),
since then any polynomial in $x_1,\ldots, x_k$ is a generalized permanent,
and a polynomial function of  the traces of  $x_i$'s means any
polynomial in $x_1,\ldots,x_k$. This, i.e., the 
general $\#P \not = NC$ conjecture in characteristic zero, 
cannot be proved unconditionally at present.
But the next case of this implication,  $n>1$ and $k$ arbitrary, can be:

\vfil \eject 
\begin{theorem}

\noindent {\bf (A mathematical form of the  $\#P \not = NC$ conjecture)}
 \label{tgenweaklec1}

The implication above holds unconditionally for any $n>1$ and arbitrary $k$.

In fact, something stronger then holds. Namely, 
when  $n>1$ and $k$ is arbitrary, no generalized permanent $p(X_1,\ldots,X_k)$
can be expressed as a polynomial function of the traces of 
$\bar X_i^j$, $1 \le i \le k$, $j \ge 0$,  where 
$\bar X_i=B_i X_i C_i$,  $i \le k$,  for any $n\times n$ complex (possibly
singular) matrices $B_i$ and $C_i$.
\end{theorem}

When  $k=1$ and $p(X_1)$ is the usual permanent, this specializes to
Proposition~\ref{pweaklec1}.

We are calling this a mathematical form for two reasons. 
First, it needs  no restriction on the computational complexity of
$p(X_1,\ldots,X_k)$ or the polynomial in the traces,
(though for trivial reasons
we can assume without loss of generality that 
the polynomial in the traces is computable in $2^{\poly(n,k,d)}$ time, 
where $d$ is  the degree of $p$). 
Thus it is rather in the spirit of the classical result of Galois theory 
which says that a  polynomial  whose Galois group is not
solvable cannot be solved by any number of
radical operations, without any restriction on the number of such
operations (though again there is a trivial  upper bound on the number of
such operations needed if the polynomial is solvable by radicals). 
Second, observe that 
the permanent has two characteristic properties: 1) the property 
P (mathematical), and 2) $\#P$-completeness (complexity-theoretic). 
The usual complexity theoretic form of the $\#P\not = NC$ conjecture
is a lower bound for  all polynomial functions with 
the  $\#P$-completeness property, whereas the mathematical form 
is  a lower bound for all polynomial functions with the 
property (P).  In other words,  the complexity theoretic form
is associated with the $\#P$-completeness property of the
permanent and the mathematical form with 
the mathematical property (P).

The result indicates that there is thus a chasm
between the two adjacent cases: $n=1$, $k$ arbitrary (the usual nonuniform
complexity theoretic $\#P \not = NC$ conjecture), and 
$n=2$, $k$ arbitrary (its mathematical form above). 
The complexity theoretic form is  far far  harder than the mathematical
form.

For some specific
generalized permanents (cf. Section~\ref{sweak}), this result again has
an elementary linear algebraic  proof as in \cite{boris}.
It also has an elementary linear algebraic  proof for a generic 
generalized permanent.
A more  nontrivial part of this 
result (which does not have a linear algebraic proof) 
is that it holds for any generalized permanent.
Indeed the basic difference between the complexity theoretic 
and the mathematical  settings is
the following. The complexity theoretic  (i.e. the usual) $\#P \not = NC$
conjecture 
is complete in the sense 
that if it is proved for one $\#P$-complete function (say
the permanent), it automatically 
holds for all $\#P$-completeness functions (because
of the theory of $\#P$-completeness). But there is no such completeness
theory at the mathematical level. Hence, a mathematical
lower bound  for a specific 
generalized permanent, e.g., the permanent,
does not say anything about all (even $\#P$-complete)
generalized permanents. To get 
similar completeness,   the  mathematical
form of the $\#P\not = NC$ conjecture above  covers up this lack of 
completeness theory
at the mathematical level by proving a result for all polynomial functions
with property (P), not just a specific one.

Theorem~\ref{tgenweaklec1} has two proofs through geometric invariant theory 
\cite{mumford}. The first proof uses only basic geometric invariant 
theory. Basically the proof for Proposition~\ref{pweaklec1} (or rather 
its nonhomogeneous
form) in \cite{GCT2append} works here as well; Bharat Adsul \cite{bharat}
has also independenly found a similar proof. But this  proof is naturalizable;
i.e. it cannot cross the natural proof barrier in \cite{rudich} as 
pointed out in \cite{GCT2append}. 

The second proof sketched in this article  is not
naturalizable and has a deeper structure that is crucial for
further progress in GCT.
Specifically, it uses the same proof strategy as for the general 
permanent vs. determinant problem and hence serves as a test case of 
the general proof strategy in a nontrivial special case.
Hence we shall only focus on the second proof in this article.

\section{From nonexistence to existence}
The rest of this lecture  outlines  the GCT approach to 
the general  permanent vs. determinant problem,
and then  points out 
the crucial steps in this plan which can be completely executed
for the   mathematical form  of the $\#P\not = NC$ conjecture above,
but which are conjectural at present for 
the general (i.e. complexity theoretic) permanent vs. determinant problem. 

\begin{figure} 
\begin{center}\psfragscanon
\psfrag{on}{$\{O_n\}$}
\epsfig{file=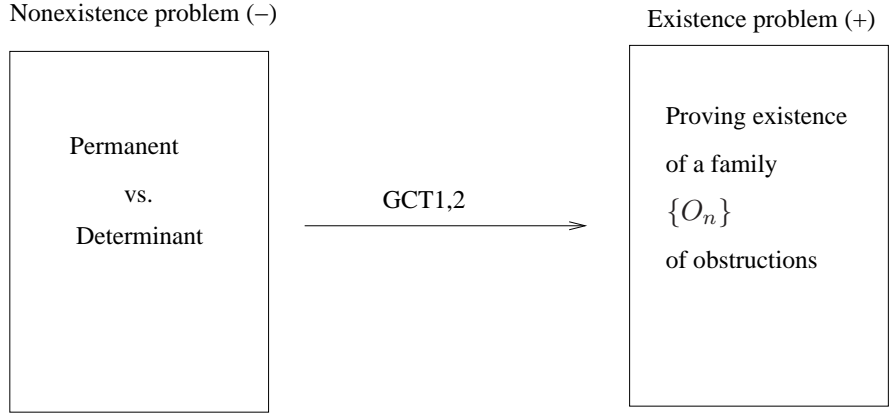, scale=.8}
\end{center}
      \caption{Reduction from nonexistence to existence}
      \label{fig:firststep}
\end{figure}

The first  step 
(GCT1,2) is to reduce this nonexistence problem--i.e., that there is no
small linear representation  of the permanent as a determinant--to
an existence problem--specifically, to the problem of proving 
existence of a family $\{O_n\}$ of obstructions; 
cf. Figure~\ref{fig:firststep}.
Here an {\em obstruction} $O_n$ is a proof-certificate of
hardness of $\perm(X)$, $X$ an $n\times n$ variable matrix,
just as the Kurotowski minor is a proof-certificate of the nonplanarity of a 
graph.  Specifically,  it is some 
algebro-geometric-representation-theoretic gadget
whose existence for every $n$  serves as a 
guarantee that $\perm(X)$ cannot be represented linearly as $\det(Y)$,
when $m=2^{\log^a n}$, $a>0$ fixed, $n\rightarrow \infty$ (i.e., for
$n$ greater than a large enough constant depending on $a$).

This reduction to existence is carried out as follows (cf. lecture 2 for 
details).

First, we associate (GCT1) with 
the complexity class $\#P$ a family $\{X_{\#P}(n,m)\}$ of (group-theoretic)
class varieties (what this means is explained below),
and with the complexity class $NC$ a family of 
$\{X_{NC}(n,m)\}$ of (group-theoretic) class varieties such that: 
if $\perm(X)$, $\dim(X)=n$, can be represented linearly 
as $\det(Y)$, $\dim(Y)=m>n$, then 

\begin{equation} \label{eqembed} 
X_{\#P}(n,m) \subseteq X_{NC}(n,m).
\end{equation} 

\begin{figure} 
\begin{center} \psfragscanon
\psfrag{xsharp}{$X_{\#P}(n,m)$}
      \psfrag{xnc}{$X_{NC}(n,m)$}
      \psfrag{g}{$G=GL_l(\C)$, $l=m^2$}
      \psfrag{Obstruction}{\mbox{Obstruction}}
\epsfig{file=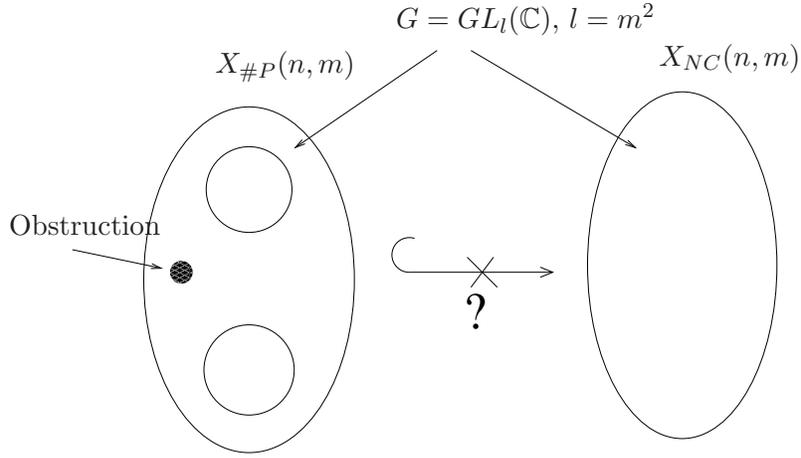, scale=.5}
\end{center}
      \caption{Class varieties}
      \label{fig:class}
\end{figure}

Each  class variety is a  (projective) algebraic variety, by which
we mean that it is the zero set of  a system of multivariate 
homogeneous polynomials with coefficients in $\C$  (akin to the 
usual curves  and surfaces). It is  group-theoretic in the 
sense that it is  constructed using group-theoretic operations 
and the general linear group $G=GL_l(\C)$, $l=m^2$, of $l\times l$ invertible
complex matrices acts on it, and furthermore the groups of
symmetries of the permanent and the determinant, which we shall
refer to as $G_{perm}$ and $G_{det}$,  are embedded in this group $G$
as its subgroups in some way. Here action means moving the points 
of the class variety around, just as $G$ moves the points of $\C^l$ around 
by the standard action via invertible linear transformations.
The goal is to show that the inclusion (\ref{eqembed}) is impossible
(obstructions are meant to ensure this);
cf. Figure~\ref{fig:class}. 

Since each class variety has a $G$-action, the space of polynomial functions 
on each class variety has a  representation-theoretic structure, which 
puts constraints on which  representations of $G$ can live on that variety
(i.e., in the space of polynomial  functions on that variety).
Informally, an obstruction is an irreducible (minimal)  representation of $G$
that can live on $X_{\#P}(n,m)$ but not on $X_{NC}(n,m)$; 
cf. Figure~\ref{fig:class}.
Existence of  an obstruction $O_n$, for every $n$, assuming
$m=2^{\log^a n}$, $a>1$ fixed, implies  that the inclusion (\ref{eqembed}) 
is not possible, since  $O_n$ cannot live  on $X_{NC}(n,m)$.
Thus  an obstruction blocks the inclusion (\ref{eqembed}).

To define an obstruction formally, we need to recall some basic representation
theory.  By a classical result of Weyl, the irreducible 
(polynomial) representations of $G=GL_l(\C)$ are in one-to-one correspondence 
with the partitions  $\lambda$ of length  at most $l$, by which we mean 
integral sequences $\lambda_1 \ge 
\lambda_2 \cdots \ge \lambda_k > 0$, $k \le l$,
where  $k$ is called the  length
of $\lambda$. The irreducible 
representation of $G$ in correspondence with $\lambda$ is denoted 
by $V_\lambda(G)$, and is called the {\em Weyl module} of $G$ indexed by 
$\lambda$. Symbolically:

\[\mbox{Irreducible representations of } G \stackrel{\mbox{Weyl}}{\Longleftrightarrow} \quad \mbox{partitions} \ \lambda. \]  

\[\mbox{Weyl module  } V_\lambda(G)  \longleftrightarrow \lambda. \] 
Weyl also proved that every finite dimensional representation of $G$ 
can be decomposed into irreducible representations--i.e.,
can be written as a direct sum of Weyl modules. Thus Weyl modules 
are the basic building blocks of the representation theory of 
$G$, and every finite dimensional representation of $G$ can be thought
of as a complex building made out of these blocks.

Now suppose $m=2^{\log^a n}$, $a>1$ fixed, $n\rightarrow \infty$. 
Suppose to the contrary that 

\begin{equation} \label{eqembed2}   X_{\#P}(n,m) \subseteq X_{NC}(n,m).
\end{equation}

Let $R_{\#P}(n,m)$ denote the homogeneous coordinate ring of 
$X_{\#P}(n,m)$; i.e., the ring of polynomial functions 
\footnote{Though the  functions here are not functions in usual sense;
but let us  not worry about this} on $X_{\#P}(n.m)$. 
Let $R_{\#P}(n,m)_d$ be the degree-$d$-component of $R_{\#P}(n, m)$ consisting
of functions of degree $d$. We define $R_{NC}(n,m)$ and $R_{NC}(n,m)_d$
similarly. Since $X_{\#P}(n,m)$ has the action of $G$, $R_{\#P}(n,m)$ 
also has an action of $G$; i.e., it is a representation of $G$. Hence,
$R_{\#P}(n,m)_d$ is a finite dimensional representation of $G$. 
Similarly, $R_{NC}(n,m)$ is a representation of $G$, and $R_{NC}(n,m)_d$
a finite dimensional representation of $G$.

If (\ref{eqembed2}) holds, then we get a natural map from $R_{NC}(n,m)$ to
$R_{\#P}(n,m)$ obtained by restricting a function on $X_{NC}(n,m)$ to
$X_{\#P}(n,m)$. 
By basic algebraic geometry, this map
is surjective and is a $G$-homomorphism.
Furthermore, it is degree-preserving. This means there
is a surjective $G$-homomorphism from $R_{NC}(n,m)_d$ to 
$R_{\#P}(n,m)_d$. Symbolically: 

\begin{equation} \label{eqsurjection}
R_{\#}(n,m)_d \leftarrow  R_{NC}(n,m)_d.
\end{equation}

\ignore{
\begin{figure} 
\begin{center} 
\psfragscanon
\psfrag{Xp}{\small $X_{\#P}(n,m)$}
\psfrag{Xnc}{\small $X_{NC}(n,m)$}
\epsfig{file=restriction.eps, scale=.6}
\end{center}
      \caption{Restriction}
      \label{fig:restriction}
\end{figure}
}

Let $R_{\#P}(n,m)_d^*$ denote the dual of $R_{\#P}(n,m)_d$; i.e., the 
set of linear maps from $R_{\#P}(n,m)_d$ to $\C$. Then (\ref{eqsurjection})
implies that there is an injective $G$-homomorphism from 
$R_{\#P}(n,m)^*_d$ to $R_{NC}(n,m)^*_d$. Symbolically: 

\begin{equation} \label{eqinclusion}
R_{\#}(n,m)_d^*  \hookrightarrow {NC}(n,m)_d^*.
\end{equation}

\begin{defn} (GCT2)
An obstruction $O_n$ is a Weyl module $V_\lambda(G)$ that 
occurs as a subrepresentation 
in $R_{\#P}(n,m)_d^*$  but not in $R_{NC}(n,m)_d^*$, for some $d$. 
We call $\lambda$ an {\em obstruction label}, and sometimes,
by abuse of notation, an obstruction as well.

A strong obstruction $O_n$ is a Weyl module $V_\lambda(G)$ 
that occurs as a subrepresentation in $R_{\#P}(n,m)_d^*$ but 
does not contain a nonzero 
invariant (fix point) of the subgroup $G_{det} \subset G$ of the symmetries 
of the determinant.
It can be shown \cite{GCT2} that a strong obstruction
is an obstruction in the above sense.
\end{defn}

Here by  an invariant we mean a point in $V_{\lambda}(G)$ 
which is fixed (does not move) with respect to the action of the subgroup
$G_{det} \subset G$.

\begin{prop} (GCT2)
Existence of an obstruction $O_n$, for all $n\rightarrow \infty$, 
with $m=2^{\log^a n}$, $a>1$ fixed, implies $\perm(X)$, $\dim(X)=n$, 
cannot be represented linearly as 
$\det(Y)$, $\dim(Y)=m$. 
\end{prop}

This   follows just from the definition of an obstruction, and 
leads to:

\begin{goal} \label{gobs} (GCT2)
Prove existence of a (strong) 
obstruction family $\{O_n=V_{\lambda_n}(G)\}$ using 
the  exceptional 
nature of $\perm(X)$ and $\det(Y)$, i.e., using the 
properties (P) and (D) in  Section~\ref{scharsym}.
\end{goal} 

\section{Obstructions for the mathematical form} \label{sobsweak}

The following result achieves this goal for the mathematical form
(Theorem~\ref{tgenweaklec1}).
\begin{theorem}  \label{tmathobs}
There exists a (strong)  obstruction family $\{O_n\}$ for the mathematical
form  of the $\#P \not = NC$ conjecture.
\end{theorem}

This implies Theorem~\ref{tgenweaklec1}.
The notion of obstructions here is similar 
to the one in the general case. 

The proof of this result  based on the results of GCT1 and 2 in
geometric invariant theory \cite{mumford} is outlined in the third lecture.
It produces  a 
family $\{O_n=V_\lambda(G)\}$ of (strong) obstructions, with a different $G$
than in the general complexity theoretic case.
Furthermore this family  is {\em strongly explicit} in the sense  that 
the specification $\lambda_n$ of each $O_n$
has polynomial, in fact,  $O(n)$ bitlength, and  
can be constructed in polynomial, in fact, $O(n)$ time (regardless of the 
complexity  of the polynomial in the traces in the statement of 
Theorem~\ref{tgenweaklec1}).

\section{Towards existence of obstructions in general via positivity}
We now proceed to describe the main results of GCT for the general
(complexity-theoretic)
permanent vs. determinant problem in the context of
Goal~\ref{gobs}.

\begin{figure} 
\begin{center} 
\psfragscanon
\psfrag{l}{$\lambda$}
\psfrag{2l}{$2 \lambda$}
\psfrag{3l}{$3 \lambda$}
\psfrag{v1}{$V_\lambda(G)$}
\psfrag{v2}{$V_{2 \lambda}(G)$}
\psfrag{v3}{$V_{3\lambda}(G)$}
\epsfig{file=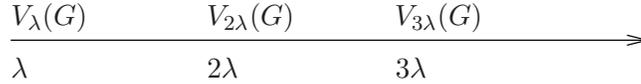, scale=1}
\end{center}
      \caption{statistics}
      \label{fig:statistics}
\end{figure}

Towards that end, we  define 
certain representation-theoretic {\em stretching} functions. 
Let $F_{\lambda,n,m}(k)$ denote the number of occurences (multiplicity)  of
the Weyl module $V_{k\lambda}(G)$ in $R_{\#P}(n,m)^*$ (i.e. 
$R_{\#P}(n,m)^*_d$, for some $d$)  as a subrepresentation.
Let $G_{\lambda,m}(k)$ denote the multiplicity of the trivial one dimensional
representation (invariant) of $G_{det} \subset G$ in 
$V_{k\lambda}(G)$. In other words, $G_{\lambda,m}(k)$ is the dimension
of the subspace of invariants of $G_{det}$ in $V_{k \lambda}(G)$.
These are statistical functions
associated with the class variety $X_{\#P}(n,m)$ 
and the subgroup embedding  $G_{det} \hookrightarrow  G$.
In the first case,  the statistics associates 
with every number (stretching parameter) $k$  
the multiplicity of the corresponding
Weyl module $V_{k\lambda}(G)$ in $R_{\#P}(n,m)^*$ and in the second 
case the dimension of the subspace of invariants of the symmetries of the
determinant in $V_{k \lambda}(G)$; cf. Figure~\ref{fig:statistics}.

Let us  call a function $f(k)$ {\em quasipolynomial}, if there exist 
$l$ polynomials $f_i(k)$, $1\le i \le l$, for some $l$, such that
$f(k)=f_i(k)$ for all nonnegative integral $k=i$ modulo $l$;
here  $l$ is
called the period of the  quasi-polynomial. 
Thus  quasi-polynomials
are hybrids of  polynomial and periodic functions. 
We say that $f(k)$ is an 
{\em asymptotic quasipolynomial} if there exist 
$l$ polynomials $f_i(k)$, $1\le i \le l$, for some $l$, such that
$f(k)=f_i(k)$ for all nonnegative integral $k=i$ modulo $l$ for 
$k\ge a(f)$, for 
some nonnegative integer depending on $f$. The minimum 
$a(f)$ for which this holds is called the {\em deviation from 
quasipolynomiality}. Thus $f(k)$ is a (strict) quasipolynomial when this
deviation is zero.

A fundamental example of a quasi-polynomial is the Ehrhart quasi-polynomial
$f_P(k)$ of a polytope $P$ with rational vertices. It is defined
to be the number of integer points in the dilated polytope $kP$. By the 
classical result of Ehrhart, it is known to be a quasi-polynomial.
More generally, let $P(k)$ be a polytope parametrized by nonnegative
integral $k$: i.e., 
defined by a linear system of the form: 
\begin{equation} \label{eqpara}
A x \le k b + c,
\end{equation}
where $A$ is an $m \times n$ matrix, $x$ a variable $n$-vector, and 
$b$ and $c$ some constant $m$-vectors. Let $f_P(k)$ be the number 
of integer points in $P(k)$.  It is known to be  an asymptotic
 quasi-polynomial. We shall call it the asymptotic 
Ehrhart quasi-polynomial  of the parametrized polytope $P(k)$. 
In what follows, we  denote a parametrized polytope $P(k)$ by 
just $P$. 
From the context it should be clear whether $P$ is a usual nonparametrized 
polytope or a parametrized polytope.

\begin{theorem} (GCT6) \label{tquasi}

\noindent (a) The  function $G_{\lambda,m}(k)$ is a quasi-polynomial.

\noindent (b) The function $F_{\lambda,n,m}(k)$ is 
an asymptotic quasi polynomial.

Analogous result holds for the $P$ vs. $NP$ problem in characteristic zero.
\end{theorem}

\begin{figure} 
\begin{center} 
\epsfig{file=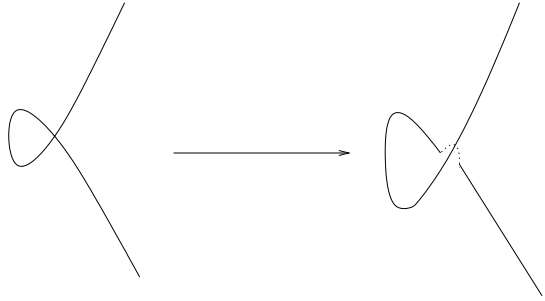, scale=.5}
      \caption{Resolution}
      \label{fig:resolution}
\end{center}
\end{figure}

The proof of Theorem~\ref{tquasi} is based on:
\begin{enumerate} 
\item The classical work of Hilbert in invariant theory, 
\item The 
resolutions of singularities in characteristic zero
\cite{hironaka}: 
this  roughly says 
that the singularities of any algebraic variety in characteristic zero
can be untangled (resolved) nicely in a systematic fashion; cf.
 Figure~\ref{fig:resolution}.
\item Cohomological works of Boutot, Brion,  Flenner, Kempf  and others
based on this resolution; cf.  \cite{boutot,dehy} and
GCT6 for the history and other references.
\end{enumerate}

As such, this proof is highly nonconstructive. It gives no effective
bound on the period $l$--it just says that $l$  is finite.

\noindent {\em Remark:} 
The original IAS lecture   stated a conditional form of (b),
which said that $F_{\lambda,n,m}(k)$ is a quasi-polynomial if the
singularities of the class variety $X_{\#P}(n,m)$ are rational and normal 
(in some algebro-geometric sense). 
Recently, Shrawan Kumar  \cite{shrawan}  has shown that 
the singularities of $X_{\#P}(n,m)$ are not normal if $m>n$. This means 
$F_{\lambda,n,m}(k)$ need not be  a quasi-polynomial and 
asymptotic quasi-polynomiality as in  (b) is all that we can expect.
This is fine 
as long as the singularities of $X_{\#P}(n,m)$ are not too bad 
in the sense described in Remark 3  after Hypothesis~\ref{hph1} below.

The following hypothesis 
says that $G_{\lambda,m}(k)$  can be realized as the Ehrhart quasi-polynomial
of a polytope, and 
$F_{\lambda,n,m}(k)$ 
can be realized as the asymptotic Ehrhart quasi-polynomial 
of a parametrized  polytope.

\begin{hypo} {\bf (PH) [Positivity Hypothesis]} \label{hph} (GCT6)

\noindent (a) For every $\lambda,n,m\ge n$, 
there exists   a parametrized  polytope $P=P_{\lambda,n,m}(k)$ 
such that 
\begin{equation} \label{eqP}
F_{\lambda,n,m}(k)=f_P(k).
\end{equation} 
It is also assumed here that there exists for $P_{\lambda,n,m}$ a specification
of the form (\ref{eqpara}),
 where $A$ is independent of $\lambda$, and $b$ and $c$ 
are  piecewise homogeneous linear functions of $\lambda$.

\noindent (b) For every $m$, there exists a (usual nonparametrized) 
polytope  $Q=Q_{\lambda,m}$  such that
\begin{equation} \label{eqQ}
G_{\lambda,m}(k)=f_Q(k).
\end{equation}
It is assumed here that there exists for $Q_{\lambda,m}$ a 
specification of the form (\ref{eqpara}) with $k=1$ and $c=0$  
where $A$ is independent of $\lambda$, and $b$ is 
is a  piecewise homogeneous linear function of $\lambda$.

Analogous positivity hypothesis also holds 
for the $P$ vs. $NP$ problem in characteristic zero.
\end{hypo}

If such $P$ and $Q$ exist, their dimensions are guaranteed to be
small by the proof of Theorem~\ref{tquasi}: specifically,
the dimension $P$ is guaranteed to be bounded by a 
polynomial in $n$, and the dimension of $Q$ 
by a polynomial in $n$ (but independent 
of $m$), if the length of $\lambda$ is $\poly(n)$ (as it would be in 
our applications).

When PH holds, we say that $F_{\lambda,n,m}(k)$ and 
$G_{\lambda,m}(k)$ have  {\em positive convex representations}. 
Here positivity refers to the fact that the Ehhart function 
$f_P(k)$ is a positive expression:
\[ f_P(k)=\sum_{v} 1,\] 
where $v$ ranges over all integer points in $P(k)$--there are no 
alternating signs in this expression.
Convexity refers to the  convexity of the polytopes $P(k)$ and $Q$.

But, a  priori, it is not at all clear why PH should  even hold. 
Many numerical functions in mathematics are quasi-polynomials 
or asymptotic quasi-polynomials (e.g., the Hilbert function 
\footnote {Hilbert function $h_Z(k)$ of a projective algebraic variety $Z$
is defined to be $\dim(R(Z)_k)$, where $R(Z)$ is the homogeneous coordinate 
ring of $Z$ and $R(Z)_k$ its degree $k$-component.} of any projective
variety), but they rarely have positive convex representations.
PH is expected to hold 
because of the exceptional nature of the determinant and the permanent. 
For concrete  mathematical evidence and justification, see  GCT6,7, and 8.

The hypothesis 
PH alone is not sufficient to prove the existence of obstructions.
The precise statement of a sufficient condition is given in 
the theorem below.

\begin{theorem} (GCT6) \label{tgct6}
There exists a family $\{O_n\}$ of (strong) obstructions 
for the  $\#P$ vs. $NC$ problem in characteristic zero, 
for $m=2^{\log^a n}$, $a>1$ fixed, $n\rightarrow \infty$, 
assuming, 

\begin{enumerate} 
\item PH, and
\item OH (Obstruction Hypothesis): 

For all $n\rightarrow \infty$,  there exists $\lambda$ such that 
$P_{\lambda,n,m}(k) \not = \emptyset$ for all large enough $k$ and
$Q_{\lambda,m}=\emptyset$.
\end{enumerate}

Analogous result holds for the $P$ vs. $NP$ problem in characteristic zero.
\end{theorem}

Mathematical evidence and arguments in support of OH are given in GCT6.
The analogous OH that arises in the context of the mathematical form
of the $\#P\not = NC$ conjecture can be  proven unconditionally.

We call  $\lambda$ a {\em polyhedral  obstruction}
(or rather, polyhedral  obstruction-label)
if it satisfies OH. In this case, $k\lambda$, for some integer $k\ge 1$ is
a (strong)  obstruction--we just have to choose $k$ large enough 
so that 
$P_{\lambda,n,m}(k)$ contains an integer point.
Henceforth, whenever we say obstruction, we actually mean polyhedral
obstruction.

There is a fundamental difference between the nature of PH and OH.
PH is a {\em mathematical hypothesis}, because there is no 
constraint on what $m$ should be in comparison to $n$ in its statement.
In contrast, OH is a {\em complexity theoretic hypothesis}, because 
$m$ needs to be small in comparison to $n$ for it to hold.



\section{The flip: Explicit construction of obstructions} 
\label{sflip}
In principle PH may have a nonconstructive proof (like that
of Theorem~\ref{tquasi})  which only tells  that such polytopes 
exist without explicitly constructing them. But proving OH may not be feasible
unless the polytopes $P$ and $Q$ in PH are given explicitly.
This  suggests the following strategy for proving 
existence of obstructions proposed in  GCT6 and  GCTflip.

\noindent (1) Prove the following stronger explicit form 
of PH, which is reasonable since the polytopes $P$ and $Q$, if they exist,
are already guaranteed to be of small (polynomial) dimension:

\begin{hypo} {\bf (PH1)}  \label{hph1} (GCT6)

\noindent (a) There exists an {\em explicit} 
parametrized polytope $P_{\lambda,n,m}=P_{\lambda,n,m}(k)$ 
as in  PH (a).
Explicit means: 

\begin{enumerate}
\item  The polytope is specified by an explicit system of linear constraints,
where the bitlength of (the description of) 
each constraint is  $\poly(n,\bitlength{\lambda}, \bitlength{m})$.
Here and in what follows, $\bitlength{z}$ denotes the bitlength of the 
description of $z$.

\item  The membership problem for the polytope
$P_{\lambda,n,m}(k)$  also 
belongs to the complexity class $P$.
That is,  given a point $x$, whether it belongs to $P_{\lambda,n,m}(k)$ 
can also be decided in $\poly(\bitlength{x},\bitlength{\lambda},
n,\bitlength{m})$
time.
Furthermore, we assume that if $x$ does not belong to the polytope,
then the membership algorithm also gives a hyperplane separating 
$x$ from the polytope in the spirit of \cite{lovasz}. 
\end{enumerate} 

\noindent (b) There exists a similar  explicit 
(nonparametrized) polytope $Q_{\lambda,m}$  satisfying   PH (b)
with the polynomial bounds that 
depend on $n$ and the bitlength of $\lambda$, but not on $m$.
\end{hypo}

\noindent {\em Remark 1:} 
Note the occurrence of  $\bitlength{m}$  instead of $m$ in the polynomial
bounds in (a) (which implies that the bounds here become $\poly(n)$
when $m<2^n$ and $\bitlength{\lambda}=\poly(n)$), and the absence 
of $m$ in the polynomial bounds in (b), which means they again are 
$\poly(n)$ when  $\bitlength{\lambda}=\poly(n)$).
The reasons for this will be explained in Lecture 3 
(cf. remarks before Hypothesis~\ref{hph1tilde} and 
after Hypothesis~\ref{hph1restatement}).

\noindent {\em Remark 2:} 
In particular, PH1 implies that the functions $F_{\lambda,n,m}(k)$ and
$G_{\lambda,m}(k)$ belong to the complexity class $\#P$. 

\noindent {\em Remark 3:} PH1 
also implies that the deviation from quasipolynomiality 
of $F_{\lambda,n,m}(k)$ is small, specifically, 
$2^{O(\poly(\bitlength{\lambda},n,\bitlength{m}))}$, i.e.,  the bitlength
of the deviation is polynomial.  As remarked after Theorem~\ref{tquasi},
this deviation would have been zero, i.e., $F_{\lambda,n,m}(k)$ would
have been a (strict) quasi-polynomial, if the singularities of the class
variety $X_{\#P}(n,m)$ were all normal and rational, which, as we know now,
is not the case \cite{shrawan}. So small deviation from quasi-polynomiality
implied by PH1 
basically means that the deviation from rationality and normality of
the singularities
of the class variety $X_{\#P}(n,m)$ is small (cf. Theorem~\ref{tquasi}).
This is the basic minimum that is required by PH1.

Like PH, PH1 is also a mathematical hypothesis in the sense that
it puts  no constraint on what $m$ should be in comparison to $n$. 
Of course, unlike PH, there is some complexity theoretic aspect to it, 
but it is secondary in comparison to the complexity-theoretic aspect of 
OH, since smallness  of $m$ with respect to  $n$ is the 
crux of the lower bound  problems under consideration.

\noindent (2a) {\bf [The flip]}

Let $m=2^{\log ^a n}$, for a fixed $a>1$. 
Using the explicit forms of the polytopes $P$ and $Q$ in PH1,
show existence of  an {\em explicit} family $\{O_n=V_{\lambda_n}(G)\}$ 
of (polyhedral) obstructions  satisfying OH. 
We say that an obstruction (proof-certificate) 
$\lambda_n$ is {\em explicit} if is ``short'' and ``easy to verify'':

\begin{enumerate} 
\item  Short: This means its bitlength $\bitlength{\lambda_n}$ 
is $\poly(n)$,
regardless what $m$ is, as long as it is $\le 2^{\log^a n}$, for some fixed
$a>1$.  

\item Easy to verify: given $n, m \le 2^{n}$ and 
$\lambda_n$, whether 
$\lambda_n$ is a valid polyhedral obstruction 
can be verified 
in $\poly(n,\bitlength{\lambda})$ time. In particular, this is
$\poly(n)$ when $\bitlength{\lambda}=\poly(n)$. 
\end{enumerate}

Existence of an explicit family of polyhedral 
obstructions is equivalent to saying
that the problem of deciding existence of polyhedral obstructions
for given $n$ in unary and $m$ in binary 
belongs to $NP$--we shall refer to this decision problem as $DP(OH)$.
This definition of explicitness is quite natural since the class $NP$ 
is a class of problems whose proof-certificates (witnesses) are 
short and easy to verify. As such, the flip--going for 
explicit  obstructions--is
a proof-strategy that is literally given to us on a platter 
by the $P$ vs. $NP$ 
problem itself. Why it is called flip will be explained later.

It should be stressed that we are primarily interested in only proving 
existence of obstructions. Whether they are explicit or not 
does not really matter in the original statement of the problem.
But we need to know the polytopes $P$ and
$Q$ explicitly (as in PH1)  so that proving OH is feasible. 
But once  PH1 is proved, existence of an explicit family follows
automatically,   as a bonus, whether we care for it or not. 

To see why, let us observe that 
the second condition above  (ease of verification) follows directly
from PH1 and the polynomial time algorithm for 
linear programming on  polytopes given by separation oracles \cite{lovasz}. 
Shortness  also follows from PH1. 

Thus it is as if the $P$ vs. $NP$ problem is forcing us to go
for explicit obstructions.

\noindent (2b) {\bf [The strong flip (optional)]} 

Using the explicit forms of the polytopes $P$ and $Q$ in PH1,
construct (rather than just show existence of) 
a {\em strongly explicit} family $\{O_n=V_{\lambda_n}(G)\}$ 
of obstructions  satisfying OH. We say that an explicit family 
of obstructions is strongly explicit if,
for each $n$, a valid obstruction-label
$\lambda_n$ can be constructed in $\poly(n)$ time.
In particular, the height and the bitlength of
 $\lambda$ is $\poly(n)$ (short) 
regardless what $m$ is, as long as it is $\le 2^{\log^a n}$, for some fixed
$a>1$.

For the purposes of the lower bound problems that we are interested in, the
flip (just explicit  
existence) would suffice and the stronger flip (explicit construction)
is optional. But the stronger flip can give us deeper insight into these
lower bound problems; cf. \cite{GCTexplicit} and GCTflip for more on this.

Now we turn to a few obvious questions.

\section{What has been achieved by all this?: The meaning of the flip} 
\label{smeaning}
Let us now see what has been achieved so far
in the context of the 
$\#P\not = NC$ conjecture in the nonuniform setting (characteristic zero),
the argument for the $P$ vs. $NP$ problem in characteristic zero
being  similar. 
At  first glance,
it may seem that all that GCT has achieved 
is to exchange a known difficult problem of 
complexity theory with a new very difficult problem of algebraic geometry.
In order to see that something is gained in exchange let us reexamine the 
original question.

\begin{sidewaysfigure} 
\begin{center}
\psfragscanon
\psfrag{sigma2p}{{\small $\Pi_2^{\#P}$}}
\psfrag{sigma1p}{$\Sigma_1^{\#P}$}
\psfrag{P}{{\small $P$}}
\psfrag{NP}{{\small $NP$}}
\psfrag{T}{$T$}
\psfrag{#P}{$\# P$}
\psfrag{inNP}{$DP(OH) \in NP$}
\psfrag{inP}{$DP(POH) \in P$}
\psfrag{insigma2p}{$DP(IOH) \in \Pi_2^{\#P}$}
\psfrag{PHplus}{PH1}
\psfrag{PHtil}{$PH^\dagger$}
\epsfig{file=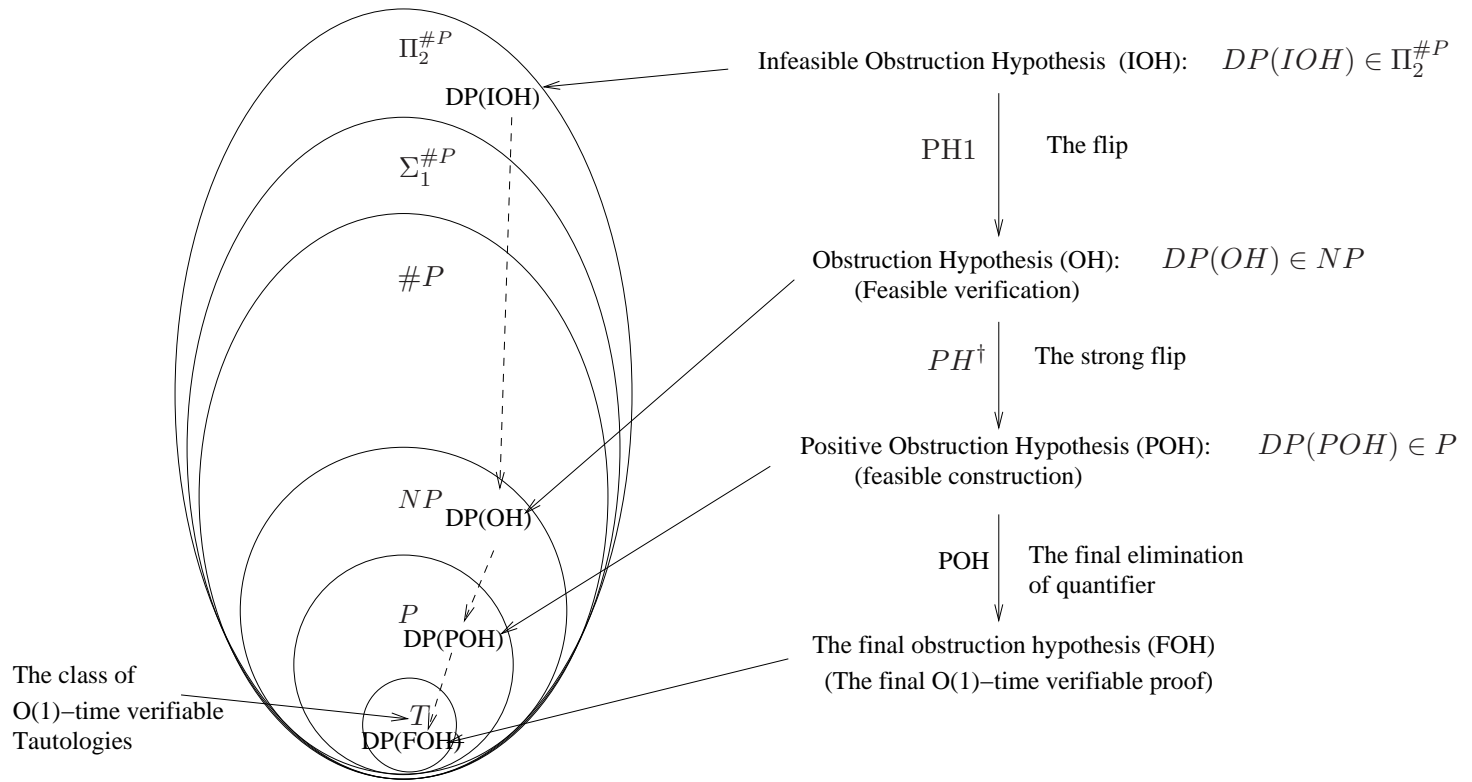, scale=.8}
\end{center}
      \caption{Positivity as a means to eliminate the quantifiers and 
reduce the complexity of the  decision problem associated 
with the obstruction hypothesis}
      \label{fig:IOHcomplexity}
\end{sidewaysfigure}

The goal of the original  conjecture is to prove that 
$\perm(X)$, $\dim(X)=n$,  cannot be computed by an arithmetic circuit $C$
of size $m=\poly(n)$, or more generally,
$m \le 2^{\log^a n}$, for some fixed $a>1$, and depth $O(\log^a n)$.
Symbolically, 
let  $f_C(X)$ denote the function computed by $C$. Then we want to prove that
\begin{equation} \label{eqioh} 
 {\bf (IOH):}  \forall n \ge n_0 \forall C \exists X:
\perm(X)\not = f_C(X),
\end{equation} 
where $n_0$ is a sufficiently large constant and $C$ ranges over 
circuits of size $m=\poly(n)$.
For given $X$ and $C$, the problem 
of deciding if  $\perm(X) \not = f_C(X)$
belongs to  $P^{\#P}$.  Let $DP(IOH)$ denote the decision problem 
of deciding for given $n$ and $m$ (in unary) whether 
(\ref{eqioh}) holds with $C$ ranging over circuits of size $m$.
Since there are two alternating layers of quantifiers in (\ref{eqioh}), it
belongs to $\Pi_2^{\#P}$, which is very high in the complexity hierarchy
(cf. Figure~\ref{fig:IOHcomplexity}).
Hence, we  refer to the original hypothesis (\ref{eqioh}) 
to be proven  as IOH 
(Infeasible Obstruction Hypothesis). Of course, IOH is expected to 
be a tautology, and hence (\ref{eqioh})
is expected to  be verifiable for small  $m=\poly(n)$ 
in $O(1)$ time--but
we do not know that as yet.

\begin{figure} 
\begin{center} 
\psfragscanon
\psfrag{eq}{$\perm(X)\not = f_C(X)$}
\epsfig{file=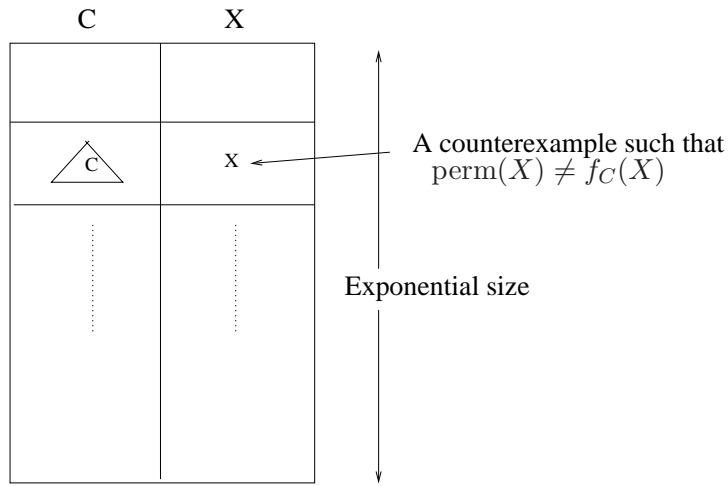, scale=.5}
\end{center}
      \caption{A trivial obstruction}
      \label{fig:trivial}
\end{figure}

Equivalently, the goal of IOH 
is to prove existence of a {\em trivial obstruction},
which is a table that lists for each small $C$
as above a counterexample  $X$ so that
$\perm(X)\not = f_C(X)$; cf. Figure~\ref{fig:trivial}.
The number of rows of this table is equal
to the number of circuits $C$'s of size $m=\poly(n)$ and depth
$O(\log^a n)$. Thus the size of this table is 
exponential; i.e., $2^{O(\poly(n))}$. (Well, only
if the underlying field of computation is finite. For infinite fields, 
such as $\Q$ or $\C$ in this paper, 
there is another notion of a trivial obstruction (cf. 
GCT6). But let us imagine that the underlying
field is finite for this argument.)
The time to verify whether a given table is a trivial obstruction
is also exponential,
and so also the time to decide if such a table exists and construct 
one (optional) for given $n$ and $m$. 
From the complexity theoretic viewpoint, this is an infeasible 
(inefficient) task. That is why we 
call  this trivial, brute force 
strategy of proving IOH, based on existence of trivial obstructions,
an {\em infeasible strategy}. 

In contrast, assuming PH1, $DP(OH)$, the decision 
problem  of deciding if a new polyhedral obstruction exists, belongs 
to $NP$ (in a stronger sense assuming that $n$ is given in unary but
$m$ is given in binary
instead of unary) as we have already observed. Thus, assuming PH1, we have 
transformed the original decision problem for trivial obstructions
$DP(IOH) \in \Pi_2^{\#P}$ to the  decision problem for the  new polyhedral 
obstructions $DP(OH) \in NP$,
in the process bringing down the time to verify an obstruction
from  exponential (for the original trivial obstruction) to polynomial
(for the new polyhedral obstruction); cf. Figure~\ref{fig:IOHcomplexity}.  
It is crucial here that PH1, the main tool for this reduction, is a 
mathematical hypothesis,  not complexity-theoretic (cf. the remarks
after Theorem~\ref{tgct6} and Hypothesis~\ref{hph1}).
The task of verifying  an obstruction has also been 
transformed from the infeasible (exponential-time) to the feasible 
(polynomial-time). Hence the name of this strategy: the flip,
from the infeasible to the feasible.
Positivity (PH1) is used in the flip as a means  to
eliminate the quantifying variables in IOH and bring down the 
complexity of the decision problem associated with the obstruction hypothesis;
cf. Figure~\ref{fig:IOHcomplexity}.

This process can be extended further. Assuming an additional
positivity hypothesis PH4 specified below,
$OH$, whose associated decision problem  $DP(OH)$ belongs to $NP$,
can be transformed 
to POH (Positivity Obstruction Hypothesis),  whose associated
decision problem $DP(POH) \in P$; i.e.,  whether a new obstruction 
exists for given $n$ in unary and $m$ in binary 
can then be decided in polynomial time,
and if so, it can also be constructed in polynomial time; 
cf. Figure~\ref{fig:IOHcomplexity}. (The hypothesis is called PH4 instead
of PH2, because PH2 and PH3 are some other positivity hypotheses in GCT6
that complement PH1). 
Once this final positivity hypothesis POH is proven, the obstruction
hypothesis is reduced to a tautology (FOH: Final Obstruction Hypothesis),
which can be verified and constructed in
$O(1)$ time--i.e., the associated decision problem $DP(FOH)$ 
is $O(1)$-time solvable. This  would then give us 
the final $O(1)$-size  proof. 

Thus the basic idea of the flip is to use positivity systematically
as a means to eliminate the quantifiers  and reduce the complexity 
of the decision problem associated with 
the  obstruction hypothesis until 
the obstruction hypothesis  is finally 
reduced to  an $O(1)$-time verifiable  tautology;
cf. Figure~\ref{fig:IOHcomplexity}. 

Now let us specify PH4 and POH.
Towards that end, let  $k \le l=m^2$ be the length of $\lambda$.
Define 

\begin{equation} 
\begin{array}{l}
\bar P_{n,m}= \{ \lambda: P_{\lambda,n,m} \not = \emptyset \} \subseteq \C^k, \\
\bar Q_{m}= \{ \lambda: Q_{\lambda,m} \not = \emptyset \} \subseteq \C^k.
\end{array}
\end{equation}

The following is a consequence of a fundamental result \cite{brion} in 
geometric invariant theory \cite{mumford}.
\begin{theorem} 
The sets $\bar P_{n,m}$ and $\bar Q_{m}$ are convex polytopes in $\C^k$.
\end{theorem}

Then: 

\begin{hypo}  {\bf PH4} (GCT6)
The membership problems for these polytopes belong to 
$P$ and so also the problem of deciding if $\mbox{vol}(\bar P_{n,m}
\setminus \bar Q_{m})$, the volume of the relative complement 
$\bar P_{n,m} \setminus \bar Q_{m}$, is nonzero (positive);
i.e. if $\bar P_{n,m} \not \subseteq \bar Q_{m}$. By polynomial time,
we mean $\poly(n,l,\bitlength{m})$ time. This is $\poly(n)$,
if $m\le 2^{o(n)}$ and $l =\poly(n)$. 
\end{hypo}

\begin{hypo} {\bf (POH)}

For all $n\rightarrow \infty$, assuming $m=2^{\log^a n}$, $a>1$ fixed, 

\[ \mbox{vol}(\bar P_{n,m} \setminus \bar Q_{m}) > 0, \] 

for $k=(n+1)^2$. 
\end{hypo} 

\ignore{
Thus POH says that  the line-shaded area in Figure~\ref{fig:POH} has positive
volume for all large enough $n$. 

\begin{figure} 
\begin{center} 
\psfragscanon
\psfrag{P}{$\bar P_{n,m}$}
\psfrag{Q}{$\bar Q_{n,m}$}
\epsfig{file=POH.eps, scale=1}
\end{center}
      \caption{POH}
      \label{fig:POH}
\end{figure}
}

\section{How to prove PH?}  \label{sphprove}
There is  a basic prototype of PH in representation theory,
which will be described in detail towards the end of Lecture 3. We shall
refer to it  as Plethysm PH. It says that the stretching functions 
akin to $F_{\lambda,n,m}(k)$ and $G_{\lambda,m}(k)$
associated with  fundamental  multiplicities 
in representation theory called plethysm constants also have analogous 
positive convex representations. 

This  is   known for a very special case of the plethysm constant 
called the  Littlewood-Richardson (LR) coefficient
$c_{\alpha,\beta}^\lambda$. It is  defined to be 
the number of occurences of the Weyl module $V_{\lambda}(G)$ in the 
tensor product of $V_{\alpha}(G)$ and $V_\beta(G)$, considered as a 
$G$-module by letting 
$G$ act on each factor of the tensor product independently. 
The classical Littlewood-Richardson rule, which we shall refer to as LR PH,
implies that the stretching function 
$\tilde c_{\alpha,\beta}^\lambda (k) = c_{k \alpha, k \beta}^{k \lambda}$
associated with the Littlewood-Richardson coefficient has a
positive convex representation.

Plethysm PH   happens to be a fundamental open problem of
representation theory, older than the $P$ vs. $NP$ problem itself. It 
has been studied intensively in the last century, and is known to be
formidable. And  now,  as explained the third lecture, it also turns out to 
be  the heart of this approach towards the $P$ vs $NP$, the
$\#P$ vs. $NC$ problems.

A basic plan to prove Plethysm PH is given in GCT6.
It is partially 
implemented in GCT7 and 8. See Figure~\ref{fig:lift} for a pictorial
depiction of the plan. It strives  to extend the  proof of LR PH based 
on the theory of the  standard quantum group \cite{drinfeld,kashiwara,lusztig}.
There it
comes out as a consequence of a  (proof of a) deep positivity result
\cite{kazhdan,lusztig}, which
we shall refer to as LR PH0. It says that   the 
tensor product of two representations of the standard quantum group 
has a canonical basis \cite{kashiwara,lusztig} 
whose structure coefficients are all positive \cite{lusztig}
polynomials (i.e., polynomials with nonnegative coefficients).
The only known proof of this
result is based on the Riemann Hypothesis over finite fields 
proved in \cite{weil2},  and the related works \cite{beilinson}. 
This Riemann Hypothesis over finite fields 
is  itself a deep positivity statement in mathematics,
from which   LR PH can  thus be 
deduced, as shown on the bottom row of Figure~\ref{fig:lift}.

If one were only interested in LR PH, one  does not 
need this  powerful machinery, because it has a  much simpler 
algebraic-combinatorial  proof. 
But the plan to extend the
proof of LR PH to Plethysm PH in GCT6,7,8 is like a huge
inductive spiral. To  make it work,  one  needs a 
stronger inductive hypothesis than Plethysm PH--this is precisely  Plethysm 
PH0 (which  will be
described in the third lecture; cf. Hypothesis~\ref{hplph0}).
Thus what is needed now 
is a systematic lifting  of the bottom arrow in Figure~\ref{fig:lift}
to the top, so as to complete the  commutative diagram, so to speak. 

Initial  steps in this direction have been taken in GCT7,8. 
First, GCT7 constructs a {\em nonstandard quantum group}, which
generalizes the notion of a standard quantum group \cite{drinfeld},
 and plays the
same role in the plethysm setting that the standard quantum group plays
in the LR setting. Second, GCT8 gives an algorithm to construct 
a canonical basis for a representation of the nonstandard quantum group
that is conjecturally correct and has the property Plethysm PH0,
which is a   generalization of LR PH0  supported by experimental
evidence.
Now what is needed to complete  the 
commutative diagram in Figure~\ref{fig:lift} is an appropriate nonstandard 
extension of the Riemann hypothesis over finite fields and the related works
\cite{beilinson,weil2,kazhdan,lusztig}
from which  Plethysm PH0 can be deduced. This--the top-right corner 
of the diagram--is the   main  open problem at the heart of this approach. 

\section{How to prove OH or POH?} 
\begin{sidewaysfigure} 
\begin{center} 
\epsfig{file=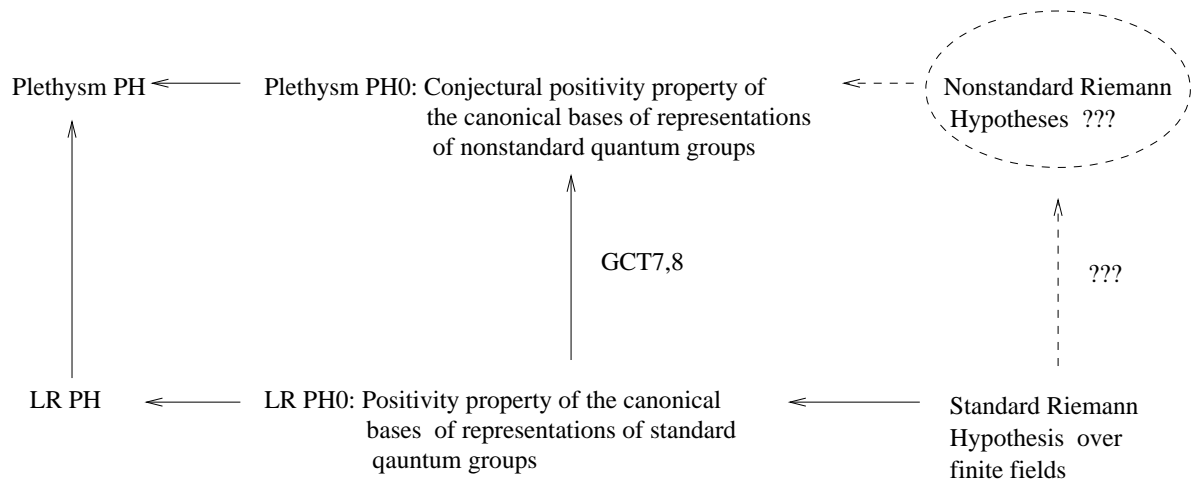, scale=.8}
\end{center}
      \caption{A commutative diagram}
      \label{fig:lift}
\end{sidewaysfigure}

We do not  know, 
since the  proof of OH and POH 
would really depend on the explicit forms of the
polytopes that arise in   PH1/PH4, and we have no idea about them
at this point.
GCT does suggest  that proving OH/POH  should be feasible 
``in theory'', i.e., {\em theoretically feasible},  assuming PH1/4,
since then DP(OH)/DP(POH)  belong to NP/P and polynomial-time
is complexity theory stands for feasible ``in theory''. 
In other words,
GCT gives a reason to believe now that proving 
the  $P\not = NP$ conjecture should be theoretically feasible. Even this
was  always questioned  in the field of complexity theory so far, because 
the $P$ vs. $NP$ problem is a universal statement regarding all of mathematics
(that says theorems cannot be proven automatically).
But, as we also know by now, 
there is a huge gap between  theory and  practice--e.g., 
just because  some problem is in $P$ does not 
necessarily mean that it is feasible in practice. Similarly,
the actual implementation of the GCT flip via positivity is expected to be 
immensely difficult ``in practice'', as Figure~\ref{fig:lift} suggests.

\section{Is positivity  necessary?} \label{sposnece}
\ignore{
Though the statement of this result is elementary, its proof is nonelementary,
as expected,  since the generalized permanents are highly nontrivial
and mathematically formidable objects. Specifically, the problem
of finding a $\#P$-basis for the space of generalized permanents 
is intimately related to some formidable problems of representation
theory and algebraic geometry,
and the  simplest analogue of this problem 
which can be solved at present  \cite{}
requires the deep Riemann Hypothesis over finite fields \cite{} for its
proof; cf. Section~\ref{}.
The proof of Theorem~\ref{} proceeds at an abstract level 
through algebraic geometry and representation theory without
having to know anything about such a basis.
At present, it is open if there is any alternative 
weaker form of the $\#P \not = NC$  conjecture,
which cannot be proved by relativizable, low-degree or 
naturalizable  techniques, but which can still be proved by elementary 
or not so  nonelementary techniques; that is, whether the earlier known
barriers mentioned above can 
be simultaneously bypassed in an elementary way in the context of 
the $\#P$ vs. $NC$ (or $P$ vs. $NC$  or $P$ vs. $NP$) problem.

\section{Why are we going for explicit obstructions?}
In a sense, the proof of the weak form turns out to be nonelementary
because of the difficulty of explicit construction. Specifically, algorithm 
to construct an obstruction $O_n$ for each $n$ is very simple, but to
show that it is correct requires nontrivial algebraic geometry,
just as the construction 
of an explicit family of Ramanujan graphs in \cite{} is very simple,
but to show that it is correct requires nontrivial algebraic geometry.
We already know, e.g. from this example of Ramanujan graphs, that 
explicit constructions can require highly nonelementary techniques.
So the question now arises: why are we going for such difficult explicit
constructions? Can they not be avoided somehow? 
After all, to solve the original problem we only need to show existence
of obstructions. We do not have to construct them explicitly.

Furthermore,  explicit construction is not needed to 
cross any of the barriers mentioned above (e.g. the natural-proof-barrier).
This is because {\em any} proof technique which uses symmetries of the
permanent would automatically bypass the natural-proof barrier
in the context of the permanent vs. determinant problem,
even if it does not utter a word about explicit construction, since
most functions have no symmetries. Indeed, the weak form of 
the generalized $\#P$ vs. $NC$ problem itself has two proofs, one
purely nonconstructive and one through explicit construction--the major 
chunk of both proofs being  the same, except for the last step,
which is carried out via a counting argument in the former proof and via
explicit construction in the latter. Since the nonconstructive proof too
is based on symmetries, it also crosses the natural-proof barrier.
This provides a concrete evidence that explicit construction is not 
necessary for crossing the earlier known  barriers.

Then why are we still going for explicit construction? 
Before we go any further, we wish to briefly state  the motivation
behind  going for the flip (explicit construction),
since this is an important issue. 
}

Finally,  if the positivity problems are so hard,  one may ask if 
they can not be avoided somehow. Unfortunately, there is a formidable 
barrier towards the $P$ vs. $NP$ and related problems, called the
{\em complexity  barrier} \cite{GCTexplicit,GCTflip}
which is universal in the sense that
any approach towards these problems would have to tackle it, not just GCT. 
The flip, i.e., explicit construction of obstructions, is the most natural
and obvious way to cross this barrier, and  the natural way may well be 
among the most effective. The 
existing mathematical evidence suggests that any such natural approach
to cross this barrier would have to say  something deep regarding positivity
(Plethysm PH/PH0)  either explicitly or {\em implicitly}, even if
the approach does not utter a word about algebraic geometry or
representation theory.
That is,  Plethysm PH and PH0   may indeed be 
the heart of the  fundamental lower bound problems in complexity theory;
a detailed story and a precise meaning of the key 
phrase {\em implicit} would appear in the revised version of GCTflip.

\ignore{\begin{figure} 
\psfragscanon
\psfrag{P}{$P\not = NP$ and $\#P\not = NC$ conjectures (characteristic zero)}
\epsfig{file=barriers.eps, scale=1}
      \caption{The feasibility barrier towards $P$ vs. $NP$}
      \label{fig:barriers}
\end{figure}}

\chapter{Class varieties and obstructions}
\begin{figure} 
\begin{center} 
\psfragscanon
\psfrag{X}{$X$}
\psfrag{Y}{$Y$}
\psfrag{n}{$n$}
\psfrag{m}{$m$}
\psfrag{z}{$z$}
\epsfig{file=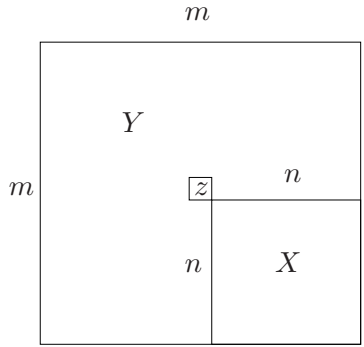, scale=.6}
\end{center}
      \caption{Variable matrix $Y$ and its submatrix $X$}
      \label{fig:matrix2}

\end{figure}

Let us begin by restating the permanent vs. determinant problem (characteristic
zero) in a form that will be convenient here.
Let $X$ be an $n\times n$ variable matrix. Let $Y$ be an $m\times m$
variable matrix, $m\ge n$. We assume that $X$ is the, say, bottom-right
minor of $Y$, and $z$ is some entry of $Y$ outside $X$, which will be 
used as a homogenizing variable; cf. Figure~\ref{fig:matrix2}.
Let $M_{m^2}(\C)$ 
denote the space of complex $m^2 \times m^2$ matrices. 
Suppose $m=2^{\log ^a n}$, $a>1$ fixed, and $n\rightarrow \infty$. 
Then the problem is
to show that there does not exist a matrix  
$A \in M_{m^2}(\C)$ such that 
\begin{equation} \label{eqpermvsdet}
\perm(X) z^{m-n} = \det(A Y),
\end{equation}
where, in the computation of $AY$, 
$Y$ is thought of as an $m^2$-vector after straightening it,
say, columnwise, and the result is brought back to the matrix form to
compute its determinant.  
It is easy to see that this problem is equivalent to the 
homogeneous restatement  of the 
permanent vs. determinant problem in the last lecture.
The best known lower bound on $m$ at present is 
quadratic \cite{ressayre}.

The  goal of this lecture:

\begin{goal} (GCT1,2)
Reduce the permanent vs. determinant 
problem to a problem in geometric invariant theory (GIT)
so that we can then start applying the machinery of algebraic 
geometry and representation theory.
\end{goal}

Specifically, 
\begin{enumerate} 
\item Define the class varieties $X_{\#P}(n,m)$ and $X_{NC}(n,m)$ 
associated with the complexity classes $\#P$ and $NC$.
\item  Define obstructions. 
\item Reduce the permanent vs. determinant problem to the problem of
showing existence of obstructions.
\end{enumerate}

For this, we need to review some basic representation theory,
algebraic geometry, and geometric invariant theory. 
The base field throughout is $\C$.

\section{Basic representation theory} \label{sbasicrepr}
Let $G$ be a group. By a representation of $G$, we mean a vector space $W$
with a homomorphism from $G$ to $GL(W)$, the space of invertible linear 
transformations of $W$. It is called irreducible if it contains no
nontrivial proper subrepresentation. 

\begin{defn} \label{dreductive}
We say that $G$ is {\em reductive} if every finite dimensional
\footnote{There are some technical restrictions on what types of
finite dimensional representations can be considered here (e.g. rational),
which we ignore here.}  representation
of $G$ is completely reducible; i.e., can be written as a direct sum
of irreducible representations.
\end{defn}

\begin{figure} 
\psfragscanon
\psfrag{x1}{$x_1$}
\psfrag{x2}{$x_2$}
\psfrag{x1=}{$x_1=x_2$}
\psfrag{x1+}{$x_1+x_2=0$}
\psfrag{Irreducible}{Irreducible subrepresentations of $S_2$}
\epsfig{file=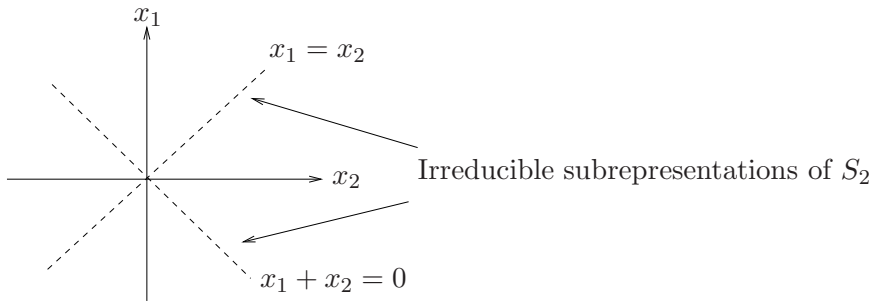, scale=.6}
      \caption{Decomposition of the standard representation of the symmetric group $S_2$}
      \label{fig:s2}

\end{figure}

All finite groups are reductive--a  classical fact \cite{fultonrepr}. 
For example, let $S_2$ be  the symmetric group on two symbols, and 
$\C^2$ its  standard representation (permutation of the coordinates). 
Then $\C^2$ is a direct sum of two irreducible subrepresentations 
given by the lines $x_1=x_2$ and $x_1+x_2=0$; cf. Figure~\ref{fig:s2}.

Weyl proved \cite{fultonrepr} 
that $G=GL_n(\C)$, the general linear group of invertible 
$n\times n$ matrices,  is  reductive,
so also  $SL_n(\C)$, the special linear group of invertible $n\times n$ 
matrices with determinant  one.

This means every finite dimensional representation $W$ of $G$ 
can be written as a direct sum:
\begin{equation} \label{eqmulti}
W=\oplus_i m_i W_i, 
\end{equation}
where $W_i$ ranges over all finite dimensional 
irreducible representations of $G$ and
$m_i$ denotes the multiplicity of $W_i$ in $W$. Thus the irreducible 
representations are the 
building blocks of any finite dimensional representation.

\begin{figure}
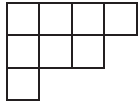
 
\begin{center}
\yng(4,3,1)
\end{center}
\caption{A Young diagram for the partition $(4,3,1)$}
\label{gyoungd}
\end{figure}

Weyl also classified these building blocks. Specifically, he showed that
the (polynomial\footnote{We say that a representation $\rho: G \rightarrow 
GL(W)$ is polynomial if the entries of $\rho(g)$, $g \in G$, are 
polynomial functions of the entries of $g$.})
irreducible representations of $G$ are in one-to-one
correspondence with the partitions (integral sequences) 
$\lambda: \lambda_1 \ge \lambda_2 \ge \lambda_k >0$ of length $k\le n$; 
we denote this partition by $\lambda=(\lambda_1,\ldots, \lambda_k)$.
It can be pictorially depicted by the corresponding
Young diagram consisting of $\lambda_i$ boxes in the $i$-th row
(Figure~\ref{gyoungd}).
An irreducible representation of $G$ in correspondence with a partition
$\lambda$ is denoted by $V_\lambda(G)$, and is called a Weyl module. 

For example, if $\lambda=(r)$, i.e., when the Young diagram consists of
just one row of $r$ boxes, then
$V_\lambda(G)$ is simply the space $\sym^r(X)$ of all homogeneous forms
of degree $r$  in
the variables $x_1,\ldots,x_n$ with the following action of $G$. Given
$f(X) \in \sym^r(X)$ and $\sigma \in G$, map $f(X)$ to
\begin{equation} \label{eqactionsym1}
f^\sigma(X) = f(X \sigma),
\end{equation}
thinking of $X=(x_1,\ldots,x_n)$ as a row vector.
This construction can be generalized to arbitrary $\lambda$ as described 
in  Appendix.

\ignore{
A Weyl module $V_\lambda(G)$ is also an irreducible representation (Weyl
module) of $\hat G=SL_n(\C)$. All irreducible  (polynomial) 
representations of $\hat G$ are obtained in this way. 
}

\section{Basic algebraic geometry} 

Let $V=\C^m$, $P(V)$ the associated projective space consisting 
of lines in $V$ through the origin, $\C[V]$ the coordinate ring of $V$,
which can also be thought of as the homogeneous coordinate ring of $P(V)$. 
Let $x_1,\ldots, x_m$ be the coordinates of $V$. 
A projective algebraic variety $Y$ in $P(V)$ is defined to be the zero set of
a set of homogeneous forms in $x_1,\ldots, x_m$ (it is also assumed 
that this zero set is irreducible; i.e., cannot be written as the union 
of two similar nonempty  zero sets). The ideal $I(Y)$ of $Y$ is defined to
be the space of all forms in $\C[V]$ that vanish on $Y$. The homogeneous 
coordinate ring $R(Y)$ of $Y$ is defined to be $\C[V]/I(Y)$.

\section{Basic geometric invariant theory} 

Now let $V$ be a finite dimensional representation of $G=GL_n(\C)$.
Then $\C[V]$ is a $G$-module (i.e., a representation) with the action 
 that, for any $\sigma \in G$,  maps  $f(v) \in \C[V]$ to 

\begin{equation} 
f^\sigma(v)= f(\sigma^{-1} v).
\end{equation} 
(This is dual of the action in (\ref{eqactionsym1})). Here $\sigma^{-1} v$
denotes $\rho(\sigma^{-1})(v)$, where $\rho: G \rightarrow GL(V)$ is 
the representation map.

\begin{defn}
A projective variety $Y \subseteq P(V)$ is called a $G$-variety if
the ideal $I(Y)$ is a $G$-submodule (i.e., a $G$-subrepresentation) of $\C[V]$.
\end{defn}

\begin{figure} 
\begin{center} 
\psfragscanon
\psfrag{P(V)}{$P(V)$}
\psfrag{G}{$G$}
\psfrag{Y}{$Y$}
\epsfig{file=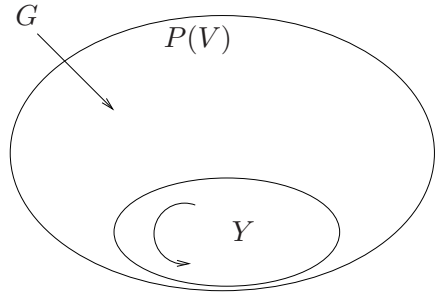, scale=.6}
\end{center}
      \caption{A $G$-subvariety of $P(V)$}
      \label{fig:gvariety}
\end{figure}

This means, under the action of $G$,
the points of $Y$ are moved to  the points within $Y$, i.e.,
each $\sigma \in G$ induces an automorphism of $Y$; cf. 
Figure~\ref{fig:gvariety}.

Let $v \in P(V)$ be a point, and $G v$ the orbit of $v$:

\begin{equation} 
G v = \{ g v \ | \ g \in G \}. 
\end{equation}

The orbit closure of $v$ is:

\[ \Delta_V[v]= \overline {G v} \subseteq P(V). \] 

The closure is taken in the complex topology on $P(V)$ by adding all limit
points of the orbit.

\begin{figure} 
\begin{center} 
\psfragscanon
\psfrag{P(V)}{$P(V)$}
\psfrag{Gv}{Orbit $G v$}
\psfrag{v}{$v$}
\psfrag{Points}{Limit points of the orbit}
\psfrag{Orbitclosure}{Orbit closure $\Delta_V[v]$}
\epsfig{file=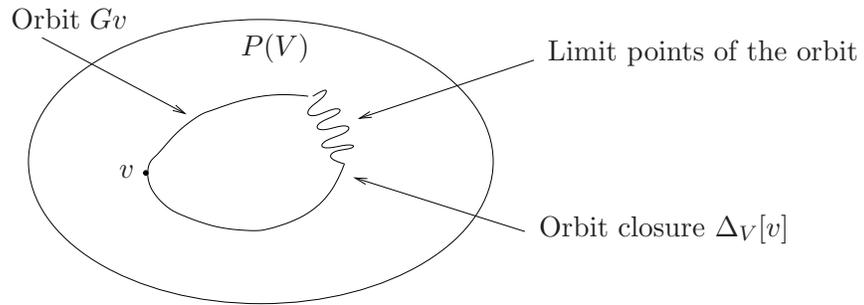, scale=.6}
\end{center}
 \caption{Orbit closure $\Delta_V[v]$}
      \label{fig:orbitclosure}
\end{figure}

\noindent Basic fact of algebraic geometry: $\Delta_V[v]$ is a 
projective $G$-variety.

The algebraic geometry of the orbit closure $\Delta_V[v]$ for general $v$ 
is  hopeless. It can be tractable only if $v$ is exceptional.

\section{Class varieties and obstructions} 
We  now construct the class varieties associated with the complexity
classes $\#P$ and $NC$ as orbit closures
of  suitable  exceptional points (the  permanent and the determinant).

Let $X,Y,z$ be as in the beginning of this lecture; cf. 
Figure~\ref{fig:matrix2}. 
Let $V=\sym^m(Y)$ be the space of homogeneous forms  of degree $m$
in the entries of
$Y$. It is a representation of $G=GL(Y)=GL_{l}(\C)$, $l=m^2$,
with the following action. Given any $\sigma \in G$, map $g(Y)$ 
to $g^\sigma(Y)=g(\sigma^{-1}(Y))$: 
\[ \sigma: g(Y) \longrightarrow g(\sigma^{-1} Y).\] 
Here $Y$ is thought of as an $m^2$-vector by straightening it,
just as in (\ref{eqpermvsdet}). 

Similarly, let $W=\sym^n(X)$ be the space of forms of degree $n$ in the
entries of $X$. It is  a representation of $H=GL(X)=GL_{n^2}(\C)$.
We define an embedding $\phi: W \hookrightarrow V$ by mapping 
any $h(X) \in W$ to $h^\phi(Y)=z^{m-n} h(X)$. This also defines an 
embedding of $P(W)$ in $P(V)$, which we denote by $\phi$ again.

Let $g=\det(Y)$. We think of it as a point in $P(V)$. 
Let $h=\perm(X) \in P(W)$. Let $f=h^\phi=\perm^\phi(Y) \in P(V)$. 
Let 
\begin{equation}
\begin{array}{lclcl} 
\Delta_V[g,m]&=&\Delta_V[g]&=&\overline{G g} \subseteq P(V), \\
\Delta_W[h,n]&=&\Delta_W[h]&=&\overline{H h} \subseteq P(W), \\
\Delta_V[f,m,n]&=&\Delta_V[f]&=&\overline{G f} \subseteq P(V). \\
\end{array}
\end{equation}
We call $\Delta_V[g,m]$ the {\em class variety}  associated with $NC$, since 
$\det(Y) \in NC$ and is $NC$-complete.
It was denoted by $X_{NC}(n,m)$ in the previous lecture;
notice that it actually depends only on $m$, and not on $n$ (the notation
was chosen to make it look symmetric like what follows). 
We call $\Delta_V[f,n,m]$ the class variety associated with $\#P$. 
It was denoted by $X_{\#P}(n,m)$ in the previous lecture. 
We call $\Delta_W[h,n]$ the base class variety associated with $\#P$.

\begin{figure} 
\begin{center} 
\psfragscanon
\psfrag{P(V)}{$P(V)$}
\psfrag{Gg}{Orbit $G g$}
\psfrag{g}{$g=\det(Y)$}
\psfrag{f}{$f=h^\phi=\perm^\phi(Y)$}
\psfrag{??}{??}
\psfrag{Points}{Limit points of the orbit}
\psfrag{Orbitclosure}{Orbit closure $\Delta_V[g]$}
\epsfig{file=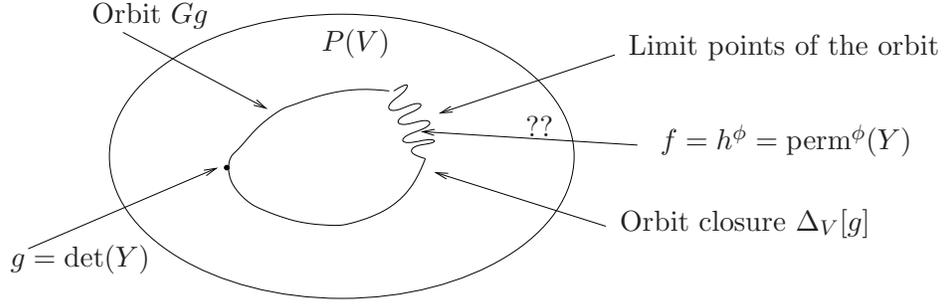, scale=.6}
\end{center}
 \caption{Does $f=h^\phi=\perm^\phi(Y) \in \Delta_V[g]$?}
      \label{fig:orbitclosure_2}
\end{figure}

\begin{prop} \label{porbitclosure} (GCT1)
If $h=\perm(X)$, $X$ an $n\times n$ matrix, can be expressed linearly 
as a determinant of an $m\times m$ matrix, $m > n$, then 
\begin{equation} 
f \in \Delta_V[g,m]=\Delta_V[g],
\end{equation}
or equivalently,
\begin{equation} 
\Delta_V[f]=\Delta_V[f,n,m] \subseteq  \Delta_V[g.m]=\Delta_V[g].
\end{equation}
Conversely, if $f \in \Delta_V[g, m]$, then $f$ can be approximated 
infinitely closely by an expression of the form
$\det(A Y)$, $A \in G$, thinking of $Y$ as an $m^2$-vector. 
\end{prop}

The first statement follows because $G=GL_{m^2}(\C)$ is dense in
$M_{m^2}(\C)$, and the second because
the $G$-orbit of $g$ is dense in $\Delta_V[g,m]$. 

\begin{conj} (GCT1) \label{cgct1orbit}
If $m=2^{\log n}$, $a>1$ fixed, $n\rightarrow \infty$, then 
$\Delta_V[f,n,m] \not \subseteq \Delta_V[g,m]$. 
\end{conj} 

By Proposition~\ref{porbitclosure},
 this would solve the permanent vs. determinant
problem in characteristic zero. 

\subsection*{How to prove the conjecture?} 

\begin{figure} 
\begin{center} 
\psfragscanon
\psfrag{Xp}{\small $\Delta_V[f]$}
\psfrag{Xnc}{\small $\Delta_V[g]$}
\epsfig{file=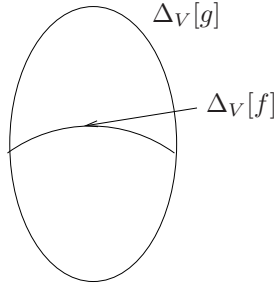, scale=.6}
\end{center}
      \caption{Restriction map}
      \label{fig:restriction2}
\end{figure}

Suppose to the contrary: 

\begin{equation} \label{eqobs1}
\Delta_V[f,n,m] = \Delta_V[f] \subseteq \Delta_V[g]=\Delta_V[g,m].
\end{equation}
Then, by basic algebraic geometry, there is a surjective homomorphism
from the homogeneous coordinate ring $R_V[g]$ of $\Delta_V[g]$ to 
the homogeneous coordinate ring $R_V[f]$ of $\Delta_V[f]$ obtained 
by restriction (Figure~\ref{fig:restriction2}).
Pictorially: 

\begin{equation} 
R_V[f,n,m] = R_V[f] \leftarrow  R_V[g]=R_V[g,m].
\end{equation}

Furthermore, since the surjection is degree preserving, 
we get a similar surjection among the degree-$d$ components: 

\begin{equation} 
R_V[f,n,m]_d = R_V[f]_d \leftarrow  R_V[g]_d=R_V[g,m]_d.
\end{equation}
Since $\Delta_V[f]$ and $\Delta_V[g]$ are $G$-varieties, these 
are (finite-dimensional)  $G$-modules. Furthermore, the homomorphism 
is a $G$-homomorphism, again by basic algebraic geometry.
By dualizing, we get an injective $G$-homomorphism from the
dual $R_V[f]^*_d$ of $R_V[f]_d$ to that of $R_V[g]_d$:

\begin{equation} 
R_V[f,n,m]_d^* = R_V[f]_d^* \hookrightarrow   R_V[g]_d^*=R_V[g,m]_d^*.
\end{equation}

\begin{defn} \label{dobstru} (GCT2)
A Weyl module $S=V_\lambda(G)$ is called an {\em obstruction} 
 for the inclusion (\ref{eqobs1}), or an obstruction for the pair $(f,g)$, 
if $V_\lambda(G)$ occurs as a $G$-submodule in $R_V[f,n,m]_d^*$ but
not in $R_V[g,m]_d^*$, for some $d$. We call $\lambda$ an 
{\em obstruction label}, and sometimes, simply an obstruction as well.
\end{defn} 

Here occurs means the multiplicity of $V_\lambda(G)$ in $R_V[f,n,m]_d^*$ 
is nonzero (cf. eq. (\ref{eqmulti})). 

If an obstruction exists for the pair $(f,g)$, for given $n$ and $m$, then
the inclusion  (\ref{eqobs1}) is not possible.
So the strategy to prove Conjecture~\ref{cgct1orbit} is to prove existence 
of such obstructions when $m$ is not too large.

\section{Why should obstructions exist?} 
But, a priori, it is not at all clear why such obstructions should even exist. 
They are expected to exist only because the 
class varieties $\Delta_V[f]$, $f=h^\phi$,  and $\Delta_V[g]$ are exceptional,
since  $h=\perm(X)$ and $g=\det(Y)$ are exceptional (cf.
Section~\ref{scharsym}).
Next, we wish to  describe in what sense the class varieties are exceptional.

\subsection{Exceptional orbit closures (group-theoretic varieties)} 
\label{sexceptional}
For that, we need to introduce the general notion of exceptional orbit 
closures.

Let $V$ be any finite dimensional representation of $G$,
$v \in P(V)$ a point, and $\hat v$ any nonzero point on the line in
$V$ corresponding to $v$. Let $H=G_{\hat v}$ be the stabilizer of $\hat v$:

\[ H= G_{\hat v} := \{ \sigma \in G \ | \ \sigma \hat v = \hat v\}. \] 

\begin{defn} (GCT1)
We say that $v$ is characterized by its stabilizer $H$ if $\hat v$ is the
only point (fix point) in $V$ such that $h v = v$ for all $h\in H$.
\end{defn} 

\begin{obs}  \label{obstriple}
If $\hat v$ is completely characterized by its stabilizer then the 
orbit closure  $\Delta_V[v]$ is completely determined by the 
associated group triple: 

\begin{equation} \label{eqgrouptriple}
 H=G_{\hat v} \hookrightarrow G \rightarrow K=GL(V), 
\end{equation} 
where the second arrow corresponds to the representation of $G$ on $V$.
\end{obs} 

Because, once we know $K$, we know $V$ (upto dual).
And once we know the embeddings 
$G \rightarrow K$ and $H \rightarrow G$, we know 
$\hat v \in V$, it being the only fix point of $H$ in $V$. 
We call (\ref{eqgrouptriple}) 
the {\em group-triple associated with $\Delta_V[g]$}. 
We also call $H\rightarrow G$ the {\em associated primary couple},
and $G\rightarrow K$ the {\em associated secondary  couple}.

\begin{defn} 
The orbit closure $\Delta_V[v]$, when $\hat v$ is completely characterized
by its stabilizer, is called a {\em group-theoretic} variety. 
\end{defn}

Coming back to the class varieties:

\begin{prop} (GCT1) \label{pchar}

\noindent (1) The determinant $\hat g=\det(Y) \in V$, $V=\sym^m(Y)$, 
is completely characterized by its stabilizer 
$G_{\hat g} \subseteq G=GL(Y)=GL_{m^2}(\C)$. 
Hence the class variety $\Delta_V[g]$ is group-theoretic.

\noindent Similarly, $\Delta_W[h]$, $h=\perm(X) \in P(W)$, $W=\sym^n(X)$, 
and $\Delta_V[f]$, $f=h^\phi$, are group-theoretic.
\end{prop}

\proof Based on classical invariant (representation) theory. 

\noindent (1) It is known that the stabilizer of 
$\hat g= \det(Y)$ in $G=GL(Y)$ is the subgroup $G_{\hat g}$ 
generated by linear transformations of the form: 

\[ Y \rightarrow A Y^* B, \quad Y^*=Y \mbox{ or } Y^t, \quad A, B \in GL_m(\C),
\] 
with $\det(A) \det(B)=1$. Ignoring this restriction, 
the continuous part $G_{\hat g}^0$  of $G_{\hat g}$ is essentially
$GL_m(\C) \times GL_m(\C)$ embedded in $G=GL_{m^2}(\C)$ naturally:

\[ G_{\hat g}^0= 
GL_m(\C) \times GL_m(\C) \hookrightarrow GL(\C^m \otimes \C^m) = 
GL_{m^2}(\C). \] 

It follows from classical representation theory that $\hat g$ is the
only fix point of $G_{\hat g}$ in $V$. 

\noindent (2) 
The  stabilizer of 
$\hat h= \perm(X) \in \sym^n(X)$
in $H=GL(X)$ is the subgroup $H_{\hat h}$ 
generated by linear transformations of the form: 

\[ X \rightarrow \lambda X^* \mu , \quad X^*=X \mbox{ or } X^t,
\] 
where $\lambda$ and $\mu$ are either  diagonal or
permutation matrices,  with obvious constraints on the 
the product of the diagonal entries when they are diagonal.

The discrete  part $H_{\hat h}^d$  of $H_{\hat h}$ is isomporphic to
 $S_n \times S_n$, $S_n$ the symmetric group, 
embedded in $H=GL_{n^2}(\C)$ naturally:

\[ H_{\hat h}^d =S_n \times S_n  \hookrightarrow GL(\C^n \otimes \C^n) = 
GL_{n^2}(\C). \] 

Again, by classical representation theory,  $\hat h$ is the
only fix point of $H_{\hat h}$ in $W$. 

\noindent (3) 
The stabilizer $G_{\hat f}$ 
of $\hat f = \hat h^\phi \in V$ in $G=GL(Y)$ 
consists of linear transformations of the 
form $Y \rightarrow A Y$, 
thinking of  $Y$ is an $m^2$-vector in which the
$n^2$ entries of its submatrix $X$ come last,
 preceded by the entry $z \in Y \setminus X$, and $A \in GL_{m^2}(\C)$
is a matrix of the form 

\[ 
\left[
\begin{array}{ccc} 
* & 0 & 0 \\
*& * & 0 \\
* & 0 & a
\end{array} \right]
\]

with $a \in H_{\hat h} \subseteq GL(X)$ (upto a constant multiple), and 
$det(A)$ suitably restricted.  The middle $*$ here acts on $z$, 
$a$ on the $X$-part of $Y$, and the $*$'s in the first column on the 
$Y\setminus (X \cup\{z\})$ part of $Y$. 
Again $\hat f$ is the only fix point of $G_{\hat f}$. \qed

\subsection{On existence of obstructions} \label{sonexi}
The main point of Proposition~\ref{pchar} is that
the information in the class varieties is completely captured by
the associated group triples. Pictorially:

\begin{equation} 
\begin{array}{cll}
\Delta_V[g] & \cong G_{\hat g} \hookrightarrow G \hookrightarrow K=GL(V), \\
\Delta_V[f] & \cong G_{\hat f} \hookrightarrow G \hookrightarrow K, \\
\Delta_W[h] & \cong H_{\hat h} \hookrightarrow H \hookrightarrow L=GL(W), \\
\end{array}
\end{equation}
where $\cong$ denotes  equivalence at the
level of information; i.e., there is  no-information-loss. 

Furthermore, by Tannakian duality \cite{delignet},
(algebraic) groups are determined by their representations; pictorially:

\begin{equation} 
 \mbox{Tannakian duality:  } \mbox{Groups} \longleftrightarrow 
\mbox{Representations}
\end{equation} 

Thus the determinant and the permanent are encoded by the
associated group triples with no information loss, and the triples, in turn,
are encoded by the associated representation theories again with no
information loss. 
This means the algebraic geometry of the class varieties is,  in principle,
completely determined by the geometric  representation theory of
the associated group triples.
Hence the difference between the class varieties--which is 
what Conjecture~\ref{cgct1orbit} is all about--should be 
reflected as a difference between
the representation theories of the associated group triples. 
This is   why  obstructions,
which can be thought as 
representation-theoretic  ``differences'', should exist. See 
GCT2 for precise mathematical results and conjectures in the Tannakian
spirit supporting this intuition.

This  leads to:

\begin{conj} (GCT2) \label{cgct2obs}
An obstruction (label) $\lambda_n$ exists for all $n\rightarrow \infty$, 
if  $m=2^{\log ^a n}$, $a>1$ fixed. 
\end{conj} 

This implies Conjecture~\ref{cgct1orbit}.

The basic plan of GCT now is:

\begin{enumerate} 
\item Understand geometric representation theory 
of the group triples associated with the class varieties
 in depth using (nonstandard) quantum groups.
\item  Translate this understanding to understand the algebraic geometry
of the class varieties.
\item Use this understanding to find obstructions as in 
Conjecture~\ref{cgct2obs}.
\end{enumerate}

\section{The flip} 
The following is a stronger form of Conjecture~\ref{cgct2obs}:

\begin{conj} {\bf [PHflip]} (cf. GCT6 and GCTflip) \label{cphflip} 
There exists an explicit family $\{\lambda_n\}$ of obstructions (labels),
if $m=2^{\log^a n}$, $a>1$ fixed, $n\rightarrow \infty$. 
\end{conj} 

Here explicit means feasible:
i.e., short and easy to verify:

\begin{enumerate} 
\item Short: the bitlength $\bitlength{\lambda_n}$ of $\lambda_n$ is
$\poly(n)=n^b$, for some fixed $b$, regardless of what $m$ is,
as long as it is not too large as above.

\item Easy to verify: The problem of verifying obstruction-labels belongs
to $P$. That is, given $n,m$ and $\lambda_n$, whether $\lambda_n$ is 
a valid obstruction-label that can belong to the above family can be
decided in $\poly(\bitlength{\lambda_n},n)$ time, again regardless 
of what $m$ is, as long as it is not too large.
\end{enumerate} 

Here one may only consider a restricted class of obstructions (labels), and 
the verification algorithm may only verify if the given label $\lambda$ 
belongs to that restricted class  in polynomial time.
This is fine as long as such  restricted $\lambda_n$ exists for every $n$.

The conjecture suggests the following basic strategy, called 
the {\em flip} (cf. GCT6, GCTflip),  for proving existence of obstructions:

\begin{enumerate} 
\item Find an ``easy'' criterion for verifying (recognizing) an
obstruction (possibly restricted). Here easy means: 
\begin{enumerate} 
\item Easy in theory: polynomial-time, and
\item Easy in practice: usable in the next step.
\end{enumerate}
\item Use this criterion to show existence of an explicit family
of obstructions. 
\item More strongly (optional), show how to construct an 
 explicit $\lambda_n$ for each $n$ in $\poly(n)$ time;
we call such a family $\{\lambda_n\}$ a strongly explicit family of
obstructions.
\end{enumerate}

Thus the flip reduces the hard nonexistence problem to the 
``easy'' existence problem for obstructions. 

\section{The $P$-barrier} \label{spbarrier}
By divine justice, finding such ``easy'' criterion for verification
is an extremely hard problem. 

To see why, let us examine 
the basic decision problems that arise in the context of  verification
of obstructions. 

\begin{problem} [Basic decision problems]
\ 

\noindent (a) Given $\lambda,n,m$, does $V_\lambda(G)$ occur in
$R_V[f,n,m]$? 

\noindent (b) Given $\lambda,m$, does $V_\lambda(G)$ occur in
$R_V[g,m]$? 
\end{problem}

Actually, the following relaxed forms of these would suffice for our purpose:

\begin{problem} [Relaxed basic decision problems]
\ 

\noindent (a)' Given $\lambda,n,m$, does $V_{k \lambda}(G)$, for some
integer $k\ge 1$,  occur in
$R_V[f,n,m]$? If so, find one such $k$. 

\noindent (b)' Given $\lambda,m$, does $V_{k \lambda}(G)$, for some integer
$k \ge 1$  occur in $R_V[g,m]$? If so, find one such $k$.
\end{problem}

We need efficient polynomial-time algorithms for these relaxed decision
problems. 

To see the main difficulty here, observe that 
the dimension of the ambient space $P(V)$ is
\begin{equation} 
M=\dim(P(V))={m^2+m-1 \choose m-1}=\mbox{exp}(m^2), 
\end{equation}
when $V=\sym^m(Y)$, and $Y$ is $m^2 \times m^2$ variable matrix. Thus 
$M$  is
the number of monomials in $m^2$ variables of degree $m$ (minus one actually). 
Furthermore, by  a classical formula of Weyl \cite{fultonrepr}, 
\begin{equation} 
\dim(V_\lambda(G))= O(\mbox{exp}(m,\bitlength{\lambda}))= 
2^{O(m+\bitlength{\lambda}}. 
\end{equation}

Currently the best unconditional algorithms for (a), (b), (a)', or (b)',
based on general-purpose 
algorithms in algebraic geometry and representation theory
take $O(\dim(\C[V]_s))$ space,
$s=|\lambda|=\sum_i \lambda_i$ (the size of $\lambda$). This is
roughly $s^M$, i.e., exponential in $M$ and hence double exponential in $m$.
The time taken is exponential in space, and hence, triple exponential in $m$. 

We cannot expect much better using such general-purpose  algorithms,
since they all use Grobner basis algorithms, and the problem of constructing
Grobner bases is EXPSPACE-complete \cite{mayr};
here EXP means  exponential in the dimension  $M$. 

Thus to get polynomial time algorithms for (a)' and (b)', 
we have to address: 

\begin{problem} {\bf [The $P$-barrier]} (cf. GCT6, GCTflip)

Bring this running time down from triple exponential in $m$ to 
polynomial in $n$.
\end{problem}

\begin{figure} 
\psfragscanon
\psfrag{Deltav}{\small $\Delta_V[g]$}
\psfrag{(general)} {(for general $g$)}
\psfrag{Deltaf}{\small $\Delta_V[f]$}
\psfrag{(general f)} {(for general $f$)}
\psfrag{det}{$\Delta_V[g]$}
\psfrag{gd}{($g=\det$)}
\psfrag{perm}{$\Delta_V[f]$}
\psfrag{fp}{($f=\perm^\phi$)}
\psfrag{Hopeless}{Hopeless}
\psfrag{H}{(triple exponential decision time)}
\psfrag{Exceptional}{Exceptional}
\psfrag{E}{(polynomial decision time??)}
\epsfig{file=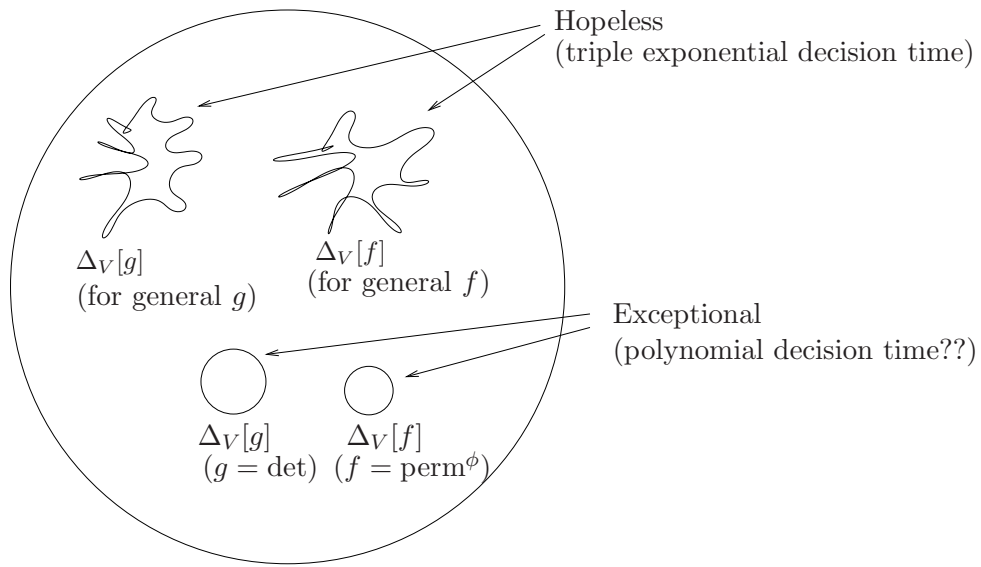, scale=.6}
      \caption{GIT chaos and the $P$-barrier}
      \label{fig:chaos}
\end{figure}

This is a massive task. 
For general $g$ and $h$, it  is impossible--i.e.,
  the problems (a)' and (b)' are  hopeless--for the reasons given
above. We refer to this as the GIT chaos (GIT=Geometric Invariant Theory);
cf. Figure~\ref{fig:chaos}. 
Conjecture~\ref{cphflip}  says, against such odds,   that this task 
should still be possible for the exceptional $g=\det(Y)$ and $h=\perm(X)$ 
that arise in GCT, and also for similar  functions
characterized by their symmetries  that arise 
in the context of the $P$ vs. $NP$ problem.

Thus the main question  here  is: 

\begin{question} 
How to cross this $P$-barrier? 
\end{question} 

GCT6 gives a plan for crossing this barrier assuming certain 
mathematical positivity hypotheses. 
This will be the subject of the  next lecture.

\chapter{Positivity}
In this lecture, we study   positivity  hypotheses in mathematics 
in the context of the problem of showing  existence of obstructions 
(Conjecture~\ref{cgct2obs}) and the $P$-barrier (Section~\ref{spbarrier}). 

Henceforth, we let 
$G=SL(Y)$ instead of $GL(Y)$ and  $H=SL(X)$ instead of $GL(X)$. This makes
no essential difference since our ambient space is $P(V)$, and two points in
$V$ differing by a nonzero scalar correspond to the same point in $P(V)$.
Thus everything discussed in the first two lectures goes through for this $G$ 
and $H$ as well.
This lecture assumes  more familiarity 
with representation theory that in the previous lectures;   Appendix
covers  the additional concepts needed here.

\section{On the $G$-module structure of the homogeneous coordinate rings
of the class varieties} \label{sgmodule}

Given $v \in P(V)$, let $\hat v$ denote any nonzero point on the line
corresponding to $v$ in $P(V)$. Let $\hat g= \det(Y) \in V$ (not $P(V)$),
and $\hat h = \perm(X) \in W$ (not $P(W))$. Let $G_{\hat g} \subseteq G$
and $H_{\hat h}$ be their stabilizers.

\begin{theorem} (GCT2) \label{tgct2first}

\noindent (1) $V_\lambda(G)$ occurs in $R_V[g]^*$ (i.e., in $R_V[g]_d^*$, for
some $d$) iff it contains a $G_{\hat g}$-invariant (i.e, a trivial 
subrepresentation--a fix point). 

\noindent (2) $V_\pi(H)$ occurs in $R_W[h]^*$ iff $V_\pi(H)$ contains an
$H_{\hat h}$-invariant. 
\end{theorem} 

This reduces some questions concerning  algebraic geometry of the class 
varieties to those concerning representation theory of the associated
group  triples (cf. Observation~\ref{obstriple} and the remarks after 
it)--or rather,
the  associated primary 
couples in this case--in keeping with the basic plan 
discussed in  Section~\ref{sonexi}.

It is easy to see that $G_{\hat g}$ is reductive (Definition~\ref{dreductive})
from its description in 
the proof of Proposition~\ref{pchar}. Hence 
$V_\lambda(G)$ contains a $G_{\hat g}$-invariant iff
the dual $V_\lambda(G)^*$ does. Thus, this theorem also holds if we replace
$V_\lambda(G)$ and $V_\pi(H)$ by $V_\lambda(G)^*$ and $V_\pi(H)^*$,
respectively.

\proof We will only prove (1), (2) being similar. 
Let $\hat \Delta_V[g,m] \subseteq V$ denote the affine cone of $\Delta_V[g,m]
\subseteq P(V)$. This is the union of all lines in $V$ corresponding to the
points in $\Delta_V[g,m]$. Thus $R_V[g,m]$, the homogeneous coordinate ring
of $\Delta_V[g,m]$, can also be thought of as the coordinate ring of 
$\hat \Delta_V[g,m]$. 

\noindent (A) [The trivial part]: Suppose $V_\lambda(G)$ occurs in
$R_V[g,m]$.  The goal is  to show that $V_\lambda(G)$ contains 
a $G_{\hat g}$-invariant.

 Fix any copy $S$ of $V_\lambda(G)$ in 
$R_V[g,m]$. 

\begin{claim} Not all functions in $S$ can vanish at $\hat g$. 
\end{claim} 

Suppose to the contrary. Then, since $S$ is a $G$-module, 
all functions in $S$ vanish on the orbit $G \hat g \subseteq V$ as well.
By homogeneity of the functions in $S$, then vanish on the 
cone of $G \hat g$ in $V$. But this cone is dense in $\hat \Delta_V[g,m]$,
since $G g$ is dense in $\Delta_V[g,m]$. 
Thus all functions in $S$ vanish on $\hat \Delta_V[g,m]$, and hence, $S$ 
cannot occur in $R_V[g,m]$; a contradiction.  This proves the claim.

Now consider the evaluation map at $\hat g$: 

\[ \psi: S \rightarrow \C,\] 
which maps every function in $S$ to its value at $\hat g$. 
It belongs to $S^*$, the dual of $S$. It is $G_{\hat g}$-invariant
since $\hat g$ is fixed by $G_{\hat g}$. 
Thus $S^*$, and hence $S$,  contains a nonzero $G_{\hat g}$-invariant. 
This proves (A).

\noindent (B) [The nontrivial part]: Suppose $V_\lambda(G)$ contains 
a $G_{\hat g}$-invariant. The goal is to show that it occurs in
$R_V[g,m]$. 

For this we need the notion of stability in geometric invariant theory
\cite{mumford},
which we now recall.

Let $Z$ be a finite dimensional $G$-representation, $G=SL_l(\C)$. 

\begin{defn} \cite{mumford} \label{dstable}
A point $z \in Z$ is called stable with respect to the $G$-action if 
the orbit $G z$ is closed in $Z$ in the complex (equivalently, Zariski) 
topology on $Z$. 
\end{defn}

\noindent Example: Let $Z=M_l(\C)$, the space of $l \times l$ complex
matrices, with the adjoint action of $G$ given by: 

\[ z \rightarrow \sigma z \sigma^{-1},\] 
for any $z \in Z$ and $\sigma \in G$. 
Then it can be shown that $z \in Z$ is stable iff $z$ is diagonalizable. 
For example, under the action of 
\[\sigma = \left[ \begin{array}{ll} t&0 \\ 0 & t^{-1} \end{array} \right], \] 
we have: 

\[ 
z=\left[ \begin{array}{ll} 1& a \\ 0 & 1 \end{array} \right]
\stackrel{\longrightarrow}{\sigma}
\left[ \begin{array}{ll} 1& a t^2  \\ 0 & 1 \end{array} \right] 
\stackrel{\longrightarrow}{t\rightarrow 0}
\left[ \begin{array}{ll} 1& 0  \\ 0 & 1 \end{array} \right].
\] 
Thus the orbit of the nondiagonalizable $z$ contains a  diagonalizable 
limit point,  which cannot be contained in the orbit. Hence  $z$ is
not stable.

Most points in any representation $Z$ of $G$ are stable \cite{mumford}.
The nontrivial
problem is  to show that a specific $z \in Z$ is stable.
For this, there is a very useful Hilbert-Mumford-Kempf criterion of
stability \cite{mumford}, using which can be proved:

\begin{theorem} (GCT1)
The point $\hat g=\det(Y) \in V=\sym^m(Y)$ is stable with respect to
the action of $G=SL(Y)$. That is, the orbit $G {\hat g} \subseteq V$ is
closed in $V$. 
\end{theorem}

Now let us get back to (B). 
Since $\hat g$ is stable, the orbit $G \hat g$ is closed in $V$,
and hence in $\hat \Delta_V[g]=\hat \Delta_V[g,m] \subseteq V$. 
That is, the orbit $G \hat g$  is a closed affine subvariety of 
$\hat \Delta_V[g,m]$. Hence, there is a surjective 
$G$-homomorphism from the coordinate ring $R_V[g,m]$ of $\hat \Delta_V[g,m]$
to the coordinate ring $\C[G \hat g]$ of $G \hat g$. 

It suffices to show that $S=V_\lambda(G)$ occurs in $\C[G \hat g]$. 
Now $G \hat g \cong G/L$, where $L=G_{\hat g}$ is stabilizer of $\hat g$.
By the algebraic form of the Peter-Weyl theorem \cite{fultonrepr}, 
the coordinate ring $\C[G]$ of $G$ (considered as an affine variety) 
decomposes as a $G$-module:
\[ \C[G] \cong \oplus_\alpha V_\alpha(G) \otimes V_\alpha(G)^*.\] 

Now 
\[ \C[G/L] = \C[G]^L,\] 
the ring of $L$-invariants in $\C[G]$. 
Thus, 

\[\C[G \hat g]= 
\C[G/L] = \C[G]^L = \oplus_\alpha V_\alpha(G) \otimes [V_\alpha(G)^*]^L.\]

Therefore $V_\alpha(G)$ occurs in $\C[G \hat g]$ iff 
$V_{\alpha}(G)^*$ contains an $L$-variant. Since $L$ is reductive, this
is so iff $V_{\alpha}(G)$ contains an $L$-invariant. 
Thus $S=V_\lambda(G)$ occurs in $\C[G \hat g]$. 

This implies (B), and proves Theorem~\ref{tgct2first}. 

We now wish to state a similar result 
for the coordinate ring $R_V[f,n,m]$, $f=h^\phi=\perm^\phi$.
For that, we need a few definitions. 
Let $W=\sym^n(X)$ and $V=\sym^m(Y)$ be as above.
Furthermore, let $\bar W=\sym^m(\bar X)$, where $\bar X$ is the 
$(n+1)\times (n+1)$ bottom-right
submatrix of $Y$ containing $X$ and $z$ in Figure~\ref{fig:matrix2}. Let 
$\bar H=SL(\bar X)$.  Thus we have 
\[ H=SL(X) \subseteq \bar H =SL(\bar X) \subseteq G=SL(Y).\] 
Let $h(X)=\perm(X)$,  and $\bar h(\bar X)= z^{m-n} h(X) \in P(\bar W)$. 
Let \[ \Delta_{\bar W}[\bar h]= \Delta_{\bar W}[\bar h,n,m] \subseteq 
P(\bar W)\] 
be the closure of the orbit $\bar H \bar h$.
Let $R_{\bar W}[\bar h]=R_{\bar W}[\bar h,n,m]$ be its homogeneous 
coordinate ring.

\begin{theorem} (GCT2) \label{tgct2second}

\noindent (a) $V_\lambda(G)$ occurs in $R_V[f,n,m]^*$ iff
the length of $\lambda$ is at most $(n+1)^2$ and $V_{\lambda}(\bar H)$ 
occurs in $R_{\bar W}[\bar h,n,m]^*$.

\noindent (b) If $V_\lambda(\bar H)$ occurs in $R_{\bar W}[\bar h,n,m]^*$, 
then it  contains as a subrepresentation
an $H$-module $V_\alpha(H)$ containing an  $H_{\hat h}$-invariant, where
$H_{\hat h} \subseteq H$ is the stabilizer of $\hat h$.

\noindent (c) Conversely, if $V_\alpha(H)$ contains an $H_{\hat h}$-invariant,
there exists a $\lambda$ lying over $\alpha$ such that 
$V_\lambda(\bar H)$ occurs in $R_{\bar W}[\bar h,n,m]^*$, and hence,
$V_\lambda(G)$ occurs in $R_V[f,n,m]^*$. Here lying over means 
(a) the length of $\lambda$ is $\le (n+1)^2$, and (b) 
$V_\alpha(H)$ occurs in $V_\lambda(\bar H)$, considered as an $H$-module
via the natural embedding $H=SL(X)  \subseteq \bar H=SL(\bar X)$. 
\end{theorem}

\section{A mathematical form of the  $\#P \not = NC$ conjecture} 
 \label{sweak}
We now  apply  GCT to prove 
Theorem~\ref{tmathobs} stated in
the first lecture.
In fact, the same proof technique yields a  more general result.

To state it, we need a few definitions.
Let $H=SL(X)=SL_{n^2}(\C)$ as before, and 
let $\tilde H=SL_n(\C) \times SL_n(\C)$ be embedded naturally in 
$SL(X)=SL(\C^n \otimes \C^n)$ (each $SL_n$ factor acts on the
corresponding $\C^n$). Thus the representation $\sym^n(X)$ for 
$H$ can also be considered to be a representation of $\tilde H$.

Let $\hat t=\trace(X^n) \in \sym^n(X)$ and
$\tilde H_{\hat t} \subseteq \tilde H$ its stabilizer.
It consists of all linear transformations of the form:

\begin{equation}  \label{eqstabtrace}
X \rightarrow A X A^{-1}, 
\end{equation} 
for all $A \in SL_{n}$. Thus $\tilde H_{\hat t} = SL_n$,  embedded 
in $SL_n \times SL_n$ naturally: 
\[ \sigma \rightarrow (\sigma, (\sigma^{-1})^t),\] 
for all $\sigma \in SL_n$. 
Let  $h=\perm \in P(\sym^n(X))$ as before,
$\hat h$  the corresponding point in $\sym^n(X)$, and 
$\tilde H_{\hat h} \subseteq \tilde H$
its stabilizer; it is esentially  the
stabilizer  described in the proof of Proposition~\ref{pchar}.

Let $W$  be any  polynomial 
representation of $\tilde H ^k= \tilde H \times \cdots \times \tilde H $
($k$ copies of $\tilde H$).
Let $w\in P(W)$ be a point, and $\hat w  \in W$ any nonzero point on the line
corresponding to $w$. 
We say that  $w$ is a {\em generalized trace-like} point
if $\hat w$  is an invariant of 
$\tilde H^k_{\hat t} = \tilde H_{\hat t} \times \cdots \times \tilde
 H_{\hat t}$ ($k$ copies of $\tilde H_{\hat t}$); i.e., 
$\tilde H^k_{\hat t} \subseteq \tilde H^k_{\hat w}$. 
We  say it is a 
{\em generalized permanent-like} point  if similarly
$\tilde H^k_{\hat w} \subseteq \tilde H^k_{\hat h}$.
We say that it is a {\em generalized
permanent} if $\tilde H^k_{\hat w} = \tilde H^k_{\hat h}$.

As an example, let $\C[X]$ be the ring  of polynomial functions in
the entries $x_{ij}$ of $X$, with the natural 
action of $\tilde H=SL_n \times SL_n$ (one factor acting on the left
and the other on the right), the case of the 
more general  ring $\C[X_1,\ldots,X_k]$ being similar.
Let $\C[X]^{\tilde H_{\hat t}} \subseteq \C[X]$ be the subring 
of the invariants of $\tilde H_{\hat t}$; i.e., 
the subring of generalized trace-like points in 
$\C[X]$. It 
is generated by $\trace(X^j)$, $j\ge 0$, by (a variant of) the first 
fundamental theorem of invariant theory \cite{fultonrepr}.
A generalized permanent in $\C[X]$ is essentially the 
same as a generalized permanent  in Definition~\ref{dgenperm}
(for $k=1$). There is a slight difference between two definitions. 
In Definition~\ref{dgenperm} we let $U_i$ and $V_i$ be any matrices in
$GL_n(\C)$, whereas here we are taking them to be in $SL_n(\C)$.
Thus as per  the definition in this section $\det(X)$ is a 
generalized permanent-like function, but not a generalized permanent.
Everything  in this section holds for a generalized
permanent in Definition~\ref{dgenperm} as well.
In what follows, we shall assume that a generalized permanent is as 
defined in this section.

Let
$\C[X]^{\tilde H_{\hat h}} \subseteq \C[X]$ be the  subring of 
invariants of $\tilde H_{\hat h}$; i.e., the subring of
generalized permanent-like functions.
By the classical result of Hilbert, it
is finitely generated.
No finite explicit set of generators for this ring is known
(unlike for the ring of generalized-trace like functions). 
But an explicit basis for this ring is known. It is as follows.
To every $n \times n$ magic square $A$ of weight $r$--i.e. 
a matrix of nonnegative
integers whose each row and column sums to $r$--assign a
{\em basic generalized permanent-like function} 
\[ p_A(X)=\sum_{A'} x_{A'},\]
where $A'$ ranges over all matrices obtained by permuting the rows 
and/or columns of $A$, and $x_{A'}=\prod_{i j} x_{ij}^{a'_{ij}}$,
$a'_{ij}$ the entries of $A'$, denotes the monomial associated with
$A'$. This is a $\#P$-computable and $\#P$-complete
function of $A$ and $X$. Furthermore,
the basic generalized permanent-like functions form a basis of 
$\C[X]^{\tilde H_{\hat h}}$.
Not all   generalized permanent-like functions are 
generalized permanents. For example, $p_A(X)$, when
every entry of $A$ is one, is not a generalized permanent, since it has more
symmetries than that of the permanent.
But most generalized permanent-like functions would be generalized 
permanents.

Now let $W$ be any   polynomial representation of $\tilde H^k$,
$\tilde H= SL_n(\C) \times SL_n(\C)$.
Given any $\sigma \in \tilde H^k$
and $w \in P(W)$, 
let $w^\sigma=\rho(\sigma) (w)$,
where $\rho: \tilde H \rightarrow GL(W)$ is the representation
map. Since this map is polynomial, $w^\sigma$ is well defined 
for any $\sigma \in (M_{n}(\C) \times M_n(\C))^k$.
Let $\tilde \Delta_W[w] \subseteq P(W)$
denote the orbit closure of $w$ with  respect
to the $\tilde H^k$ action; i.e., the closure of the orbit $\tilde H^k w$.

\begin{theorem}  \label{tweak}
Let $w$ be any generalized trace-like point in $W$, and 
and $h$ any generalized permanent in $W$. 
Then, for any $\sigma \in (M_{n}(\C) \times M_{n}(\C))^k$, $w^\sigma \not = h$.
More generally,  $\tilde \Delta_W[w]$ does not contain $h$.
\end{theorem} 

This reduces to Theorem~\ref{tgenweaklec1}  when $W=\C[X_1,\ldots,X_k]$.
The proof in \cite{GCT2append} based on basic geometric invariant
theory also works in this case. But we are more interested 
here in testing the general proof strategy of GCT based on obstructions 
in this nontrivial special case.

We  now sketch the proof of 
Theorem~\ref{tweak} based on obstructions only for $W=\C[X]$. The details for 
the general case are similar and are left to the reader.
For $W=\C[X]$, Theorem~\ref{tweak}  follows from:

\begin{theorem} 
There exists a family $\{O_n\}$ of obstructions in this case.
\end{theorem}

\proof 
Let  $h\in \C[X]$ be any generalized permanent,
and $w \in \C[X]$ any generalized trace-like point.
The class varieties $\tilde \Delta_W[w], \tilde \Delta_W[h] \subseteq
P(W)$ are now defined with respect to the $\tilde H$-action
and the obstructions are $\tilde H$-Weyl-modules defined similarly.
Let $\tilde R_W[h]$ and $\tilde R_W[w]$ be the homogeneous coordinate 
rings of $\tilde \Delta_W[w]$ and $\tilde \Delta_W[h]$.

It can be using shown using  Kempf's criterion of stability \cite{mumford}
that $\hat h$ is stable with respect to the $\tilde H$ action--the proof
of this fact is  similar to the stability related proofs in GCT1. 
Specifically, in this setting Kempf's criterion in a concrete form says
that $\hat h$ is stable if the standard irreducible 
representation $\C^n \otimes \C^n$
of $\tilde H=SL_n(\C) \times SL_m(\C)$ is also an irreducible representation of
its subgroup $\tilde H_{\hat h}$, which is easy to check.
The crucial point here is that this proof needs to know only about 
the stabilizer of $\hat h$ and nothing else.
Using stability of $\hat h$  it then follows from the general 
results in GCT2  that the analogue of
Theorem~\ref{tgct2first}  (2) holds for this $h$.

The stabilizer $\tilde H_{\hat w}$ contains 
the stabilizer $\tilde H_{\hat t} \subseteq \tilde 
H$ of $\hat t=\trace(X^n) \in \sym^n(X)$ as described in (\ref{eqstabtrace}). 

We need the following two facts.

\noindent (a) Any
irreducible $\tilde H$-module is  of the form 
$V_\alpha(SL_n) \otimes V_\beta(SL_n)$. 
By the classical Schur's lemma it contains a 
$\tilde H_{\hat t}$-invariant iff $\alpha=\beta$;.
Hence, it does not contain a $\tilde H_{\hat w}$-invariant if 
$\alpha \not = \beta$.

\noindent (b) (Cf. \cite{ariki})
An irreducible representation of $\tilde H$
of the form $1 \otimes V_{\gamma}(SL_n)$,
where $1$ stands for the trivial representation of $SL_n$ 
and $|\gamma|=2n$, contains a $\tilde H_{\hat h}$-invariant
iff $\gamma$ is even--if 
$\gamma =(\gamma_1, \gamma_2, \cdots)$, 
this means every $\gamma_i$ is divisible by $2$.
Here $|\gamma|=\sum_i \gamma_i$
denotes the size of $\gamma$.

Let $\gamma$ be any even partition with $|\gamma|=2n$.
By (b) and Theorem~\ref{tgct2first} (2)
(or rather its analogue in this case mentioned above), 
$1 \otimes V_\gamma(SL_n)$ occurs in $\tilde R_W[h]^*$.
By (a), it 
does not contain a $\tilde H_{\hat w}$-invariant.
By Theorem~\ref{tgct2first} (2) again (or rather its analogue in this case),
it cannot occur in $\tilde R_W[w]^*$. 
Therefore, $1 \otimes V_\gamma(SL_n)$ is an obstruction. \qed

The proof above shows that 
any $1 \otimes V_\gamma(SL_n)$, $|\gamma| >0$, 
which contains an $\tilde H_{\hat h}$-invariant 
is an obstruction. One can  show nonconstructively, i.e., without
using \cite{ariki}, that there is such $\gamma$ for every $n$.
This then yields a nonconstructive proof of this result (whose 
major part is the same as in the explicit proof). The proof based on
basic geometric invariant theory as in \cite{GCT2append} is also
nonconstructive.

\section{From the mathematical towards the  general complexity theoretic
form} \label{sweaktogeneral}
We now discuss what is needed to lift the proof of the mathematical
form of the $\#P \not = NC$ conjecture
to the general complexity theoretic form.
There are two issues.

(1) There is a serious leak in Theorem~\ref{tgct2second},
because there can be
several $\lambda$ lying over $\alpha$, and that result does not tell us
exactly which one of them would occur in $R_{\bar W}[\bar h,n,m]$ or 
$R_V[f,n,m]^*$, nor does it
tell us which $V_\lambda(G)$'s occur in $R_{\bar W}[\bar h,n,m]_d$
or $R_V[f,n,m]^*_d$, for a fixed
$d$. Such refined information can be obtained from 
a general positivity hypothesis 
(PH: Hypothesis~\ref{hph}) for $R_V[f,n,m]$ 
(which was  denoted  by $R_{\#P}(n,m)$ in Lecture 1).
We will discuss this  issue  in Section~\ref{sposgeneral}.

(2) To use Theorem~\ref{tgct2first} and Theorem~\ref{tgct2second}
we need an effective criteria for: 

\begin{problem} \label{pinvariant}

\noindent (a) Does $V_\lambda(G)$ contain a $G_{\hat g}$-invariant? 

\noindent (b) Does $V_\pi(H)$ contain an $H_{\hat h}$-invariant? 

\end{problem} 

These are special cases of the general subgroup restriction problem 
which we discuss next in the following section. 

\section{The subgroup restriction problem}

Let $H$ be a reductive subgroup of $G=GL(V)$, where $V$ is an explicitly
given finite dimensional representation of $H$. Symbolically:

\begin{equation}  H \stackrel{\rho} \hookrightarrow G=GL(V), 
\end{equation} 

where $\rho$ denotes the representation map. 
For example, we can have $H=GL_n(\C)$, and $V=V_\mu(H)$, 
the Weyl module of $H$. Then $\mu$ specifies the representation
map $\rho$ completely, and hence, we shall also use $\mu$ in place of 
$\rho$ in this case--called the {\em plethysm} case. Symbolically: 

\begin{equation}  \label{eqplethysm}
H \stackrel{\mu} \hookrightarrow G=GL(V), \quad V=V_\mu(H).
\end{equation}

Given any partition $\lambda$, the Weyl module $V_\lambda(G)$ of 
of $G$ can be considered an $H$-module via 
the representation map $\rho$.
 Since $H$ is reductive, 
it is completely reducible as an $H$-module:

\begin{equation} 
V_\lambda(G)= \bigoplus_\pi a^\lambda_{\pi,\rho} V_\pi(H),
\end{equation}

where $a^\lambda_{\pi,\rho}$ denotes the multiplicity of $V_\pi(H)$
in $V_\lambda(G)$. In the plethysm case, we also denote $a^\lambda_{\pi,\rho}$ 
by $a^\lambda_{\pi,\mu}$, and call it the {\em plethysm constant}.

\begin{problem} [Subgroup restriction  problem]

\noindent (1) Given partitions $\lambda,\pi$ and $\rho$,
does $V_\pi(H)$ occur as a subrepresentation of $V_\lambda(G)$? That is,
is $a^\lambda_{\pi,\rho}$  positive? 

\noindent (2) Find a good positive  formula for 
$a^\lambda_{\pi,\rho}$ akin to the usual positive formula for the permanent 
which does not have any alternating signs. What 
good and positive means would be elaborated later
(cf. Hypothesis~\ref{hphple}).
\end{problem} 

\begin{problem}[Plethysm problem]  \label{pplethysm}
The special case of the subgroup restriction problem for the
representation map (\ref{eqplethysm}), obtained by replacing $\rho$ by $\mu$.
\end{problem}

The two special cases that arise in the context of Problem~\ref{pinvariant}
are: 

\noindent (1) Let $\hat g= \det(Y) \in \sym^m(Y)$,
$G=GL(Y)=GL_{m^2}(\C)$, and $H=G_{\hat g} \subseteq G$,
the stabilizer of $\hat g$; cf. the proof of Proposition~\ref{pchar} for
its description. 
If we ignore the discrete (and torus) part of the stabilizer, then 
the subgroup restriction problem here is for the embedding:

\[ GL_m\times GL_m \hookrightarrow GL(\C^m \otimes \C^m).\] 

\noindent (2) Let $\hat h =\perm(X) \in \sym^n(X)$, and $H=GL(X)$, and
$H_{\hat h}$ the stabilizer of $\hat h$; cf. the proof of 
Proposition~\ref{pchar} for its description.
If we ignore the continuous part of the stabilizer, then 
the subgroup restriction problem here is for the embedding:

\[ S_n \times S_n \hookrightarrow GL(\C^n \otimes \C^n), \]

where $S_n$ is the symmetric group on $n$ letters.

It is a classical result of representation theory that (1) can
be reduced to the plethysm problem. 
By \cite{ariki}, (2) can also be reduced to the plethysm problem. So
the plethysm problem is the fundamental special case of the 
subgroup restriction problem that we will be interested in
(though the following results also hold for the general subgroup 
restriction problem).

\section{Littlewood-Richardson problem} 

One completely understood special case of the subgroup restriction problem
is  the Littlewood-Richardson (LR) problem. This arises when 
 $H=GL_n(\C)$ embedded  in $G=H \times H$ diagonally:

\begin{equation} \label{eqdiagonallr}
\begin{array}{lcl}
H &\rightarrow & G= H \times H \\
\sigma &\rightarrow&  (\sigma,\sigma).
\end{array}
\end{equation}

Then every irreducible representation of $G$ is of the form
$V_\alpha(H) \otimes V_\beta(H)$. Considered as an $H$-module via the 
above diagonal embedding, it
decomposes:

\[ V_\alpha(H) \otimes V_\beta(H) = \oplus_\lambda c_{\alpha,\beta}^\lambda
V_\lambda(H).\] 
The multiplicities $c_{\alpha,\beta}^\lambda$ are called 
Littlewood-Richardson coefficients. Let $\tilde c_{\alpha,\beta}^\lambda(k)
= c_{k \alpha, k \beta}^{k \lambda}$ be the associated stretching functions.

\begin{theorem}  \label{tlr}

\  

\begin{enumerate} 
\item {\bf [LR PH1]} 
There exists a polytope of $P_{\alpha,\beta}^\lambda$ with description of
$\poly(\bitlength{\alpha},\bitlength{\beta},\bitlength{\lambda})$ bitlength
such that:
\[ c_{\alpha,\beta}^\lambda= \#(P^\lambda_{\alpha,\beta}),\]  the
number of integer points in $P^\lambda_{\alpha,\beta}$, and 
\[ \tilde c_{\alpha,\beta}^\lambda (k) = c_{k \alpha,k \beta}^{k \lambda}
= \# (k P^\lambda_{\alpha,\beta}) = f_{P^\lambda_{\alpha,\beta}}(k),\] 
the Ehrhart quasipolynomial of $P^\lambda_{\alpha,\beta}$. 
This provides a good positive formula for the Littlewood-Richardson 
coefficients. 

\item {\bf [Saturation Theorem]} \cite{knutson}: 
$c_{\alpha,\beta}^\lambda \not = 0$ iff $P_{\alpha,\beta}^\lambda \not 
= \emptyset$. 

\item (GCT3,\cite{knutson}) 
Given $\alpha,\beta,\lambda$, whether $c_{\alpha,\beta}^\lambda$ is nonzero
(i.e. positive) can be decided in 
$\poly(\bitlength{\alpha},\bitlength{\beta},\bitlength{\lambda})$ time.
\end{enumerate} 
\end{theorem}

Here the third statement follows from the first two by 
a polynomial time algorithm for linear programming \cite{lovasz}. 

\section{Plethysm problem}  \label{splethysm}
Let us now focus on the plethysm problem (Problem~\ref{pplethysm}). 
Let $\tilde a_{\pi,\mu}^\lambda(k) = a_{k \pi,\mu}^{k \lambda}$ be
the stretching function associated with the plethysm constant 
$a_{\pi,\mu}^{\lambda}$.  Let 
\[ A^\lambda_{\pi,\mu}(t) = \sum_{k \ge 0} \tilde a^\lambda_{\pi,\mu}(k)\]
be the associated generating function. 
It was asked in \cite{kirillov} if it is a rational function. 
The following result shows something stronger:

\begin{theorem}  \label{tplethysmqausi} (GCT6)
The stretching function $\tilde a_{\pi,\mu}^\lambda(k)$ is 
a quasi-polynomial. 
\end{theorem}
This implies, in particular, that $A^\lambda_{\pi,\mu}$ is rational
by a standard  result of enumerative combinatorics \cite{stanleyenu}. 

The proof below is motivated by Brion's proof \cite{dehy} of quasipolynomiality
of the stretching functions associated with the Littlewood-Richardson
coefficients (of arbitrary type). 

\proof 

Let $H=GL_n(\C)$, and $V=V_\pi(H)$. Let $U \subseteq H$ be 
the subgroup of  lower triangular matrices with $1$'s on the diagonal.
Then it is known (cf.
 Appendix) that there is a unique (up to constant multiple) 
nonzero point 
$\hat v = \hat v_\pi \in V$ that is stabilized by $U$; i.e., such that
$u \hat v= \hat v$ for all $u \in U$. The point $\hat v$ is called 
the highest weight vector of $V_\pi(H)$. Let $v=v_\pi$ be the corresponding
point in $P(V)=P(V_\pi(H))$. 
Then it is known that the orbit $H v \subseteq P(V)$ is already closed.
That is, the orbit closure $\Delta_V[v]$ (with respect to the $H$ action)
is just the orbit $H v$ itself. Furthermore, by Borel-Weil \cite{fultonrepr}, 
the homogeneous coordinate ring $R_V[v]$ of $\Delta_V[v]= H v$ has the
following decomposition as an $H$-module: 

\begin{equation} \label{eqrvv}
R_V[v]=\oplus_k V_{k \pi}(H)^*,
\end{equation}

where the superscript $*$ denotes the dual. 
We can also think of $R_V[v]$ as the coordinate ring of 
$\hat \Delta_V[v]$, the affine cone of $\Delta_V[v]$. 
It is known that 
the  singularities of $\hat \Delta_V[v]$ are rational 
and normal; e.g., see \cite{smith}. 

\noindent {\em Remark:} 
By normal, we mean that for each $x \in X= \hat \Delta_V[v]$, 
there exists a  (classical)
neighbourhood  $U \subseteq X$ of $x$, such that
$U \setminus (U \cap \mbox{sing}(X))$
is connected; where $\mbox{sing}(X)$ is the subvariety of $X$ consisting
of all its singular points. Rational is much more difficult to define. 
Roughly it means the following. By Hironaka \cite{hironaka}, 
all singularities of $X$ can be resolved (untangled)--cf. 
Figure~\ref{fig:resolution}.
With each singularity of $X$, one can associate a cohomological object 
that measures the difficulty of this resolution. A singularity is called
rational if this cohomological object vanishes. This means the
singularity is sufficiently nice. 

By abuse of terminology, we say that the ring $R_V[v]$ in (\ref{eqrvv})
is normal and rational. Similarly, 
it can be shown that the ring

\begin{equation} \label{eqS}
S=\oplus_\pi V_{k \pi}(H)^* \otimes V_{k \lambda} (G)
\end{equation}

is normal and rational. (Formally, this means the singularities
of the variety, or rather the
scheme, which  can be associated with this ring, are rational
and normal.)

Let $S^H$ denote the ring of $H$-invariants in $S$: 

\[ S^H= \{ s \in S \ | \ h s = s, \quad \forall h \in H \}. \] 

By (\ref{eqS}), 

\begin{equation} 
S^H= \oplus_k [ V_{k \pi}(H)^* \otimes V_{k \lambda}(G)]^H,
\end{equation}

where the superscript $H$ on the right hand side again denotes the operation
of taking $H$-invariants. 

By a classical result of Hilbert \cite{popov},
$S^H$  is  a finitely generated ring (since $S$ is finitely generated). 
Furthermore, since $S$ is normal and rational, it follows by 
Boutot \cite{boutot} that $S^H$ is normal and rational (this is the crux 
of the argument). 

Let $h_{S^H}(k)=\dim(S^H_k)$ denote the Hilbert function of $S^H$,
where $S^H_k$ denotes the degree-$k$ component of $S^H$. 

By Schur's lemma \cite{fultonrepr}, 

\[ 
\dim ([V_{k \pi}(H)^* \otimes V_{k \lambda}(G)]^H) = a_{k \pi, \mu}^{k \lambda}
=\tilde a^\lambda_{\pi,\mu}(k),
\] 
the multiplicity of $V_{k \pi}(H)$ in $V_{k \lambda}(G)$. 
Hence, 
\[ h_{S^H}(k)= \tilde a_{\pi,\mu}^\lambda(k). \] 

By  Flenner \cite{flenner}, $h_{S^H}(k)$ 
 is a quasi-polynomial, since $S^H$ is rational and
normal. 
Thus it follows that $\tilde a_{\pi,\mu}^\lambda(k)$ is also a 
quasi-polynomial. \qed

\begin{hypo} [Plethysm PH] \label{hphple} (GCT6)

There exists a polytope of $P=P^\lambda_{\pi,\mu}$ with description of 
$\poly(\bitlength{\lambda},\bitlength{\pi},\bitlength{\mu})$ bitlength
such that 

\begin{equation} 
\tilde a_{\pi,\mu}^\lambda(k) = f_P(k),
\end{equation}
the Ehrhart quasi-polynomial of $P$. In particular,

\begin{equation} 
a_{\pi,\mu}^\lambda(k) = \#(P),
\end{equation}
the number of integer points in $P$. 

This would provide the sought good positive formula for the plethysm 
constant $a_{\pi,\mu}^\lambda$ (cf. Problem~\ref{pplethysm})
\end{hypo}

Here it is assumed that the polytope is presented by a separation 
oracle as in \cite{lovasz},
and the bitlength $\bitlength{P}$ of the description
of $P$ is defined to be $l+s$, where $l$ is the dimension of the ambient 
space in which  $P$ is defined by linear constraints,
and $s$ the maximum bitlength of any defining constraint. 
Notice that the polytope $P$ here 
depends only on $\lambda,\pi$ and $\mu$ but not on 
$H$, just like the  plethysm constant $a^\lambda_{\pi,\mu}$ itself.

\begin{theorem} \label{tlinear} (GCT6)
Assuming Plethysm PH, whether $a_{k \pi,\mu}^{k \lambda} > 0$ for some
$k \ge 1$ can be decided in $\poly(\bitlength{\lambda}, \bitlength{\pi},
\bitlength{\mu})$ time. If so, one such $k$ can also be found in
polynomial time.
\end{theorem}

\proof By linear programming \cite{lovasz}. One has to just decide if
$P^\lambda_{\pi,\mu}$ is nonempty, and if so, find a vertex $v$ of $P$ and
choose  $k$ such that $k v$ has integral coordinates. 
\qed 

\section{Positivity and the existence of obstructions in  the general case} 
\label{sposgeneral}
Now we describe how positivity can help in proving the existence of
obstructions in the general case of the $\#P$ vs. $NC$ problem.

Towards that end, first we introduce a stronger notion of obstructions.
We follow the same notation as in Section~\ref{sgmodule}. Thus 
$G=SL(Y)=SL_{m^2}(\C)$ as there. 

\begin{defn} (GCT2)
A Weyl module $V_\lambda(G)$ is called a {\em strong obstruction} 
for the pair $(f,g)$, if $V_\lambda(G)$ occurs in $R_V[f,n,m]^*$,
i.e. in $R_V[f,n,m]_d^*$ for some $d$,  but does not contain a nonzero
$G_{\hat g}$-invariant.
\end{defn}

It follows from Theorem~\ref{tgct2first} (1) that a strong obstruction is also
an obstruction as per Definition~\ref{dobstru}. 
Furthermore, by Theorem~\ref{tgct2second}, we have:

\begin{prop} \label{pgct2uni}
A Weyl module $V_\lambda(G)$ is a {\em strong obstruction} 
for the pair $(f,g)$, iff 
\begin{enumerate} 
\item The length of $\lambda$ is at most 
$(n+1)^2$,
\item  $V_\lambda(\bar H)$ occurs in $R_{\bar W}[\bar h,n,m]^*$,
i.e., in $R_{\bar W}[\bar h,n,m]_d^*$ for some $d$ 
(which has to be $|\lambda|/m$). 
\item  $V_\lambda(G)$  does not contain a nonzero
$G_{\hat g}$-invariant.
\end{enumerate}
\end{prop}

Now let $G_\lambda(k)= 
 G_{\lambda,m}(k)$ 
denote the multiplicity of the trivial representation of $G_{\hat g}$  in
$V_{k \lambda}(G)$. 

\begin{theorem} \label{tplethysm2} (GCT6)
The stretching function $G_{\lambda,m}(k)$ is a quasi-polynomial.
\end{theorem}

This is proved like Theorem~\ref{tplethysmqausi};
in fact, this is essentially its special case. 

The following is a 
precise form  Hypothesis~\ref{hph} (b).
It is essentially a special case of Plethysm PH:

\begin{hypo} [PH] \label{hphvariant} (GCT6)

There exists a polytope of $Q_{\lambda}$ 
such that 

\begin{equation} 
 G_{\lambda,m}(k) = f_{ Q_{\lambda}}(k),
\end{equation}
for every $m$.
\end{hypo}

Here  the polytope $ Q_\lambda$ does not depend on $G$ or its 
dimension $m=\dim(G)$, 
for the same reasons that  the polytope $P$ in the Plethysm PH 
does  not depend on $H$ there; cf. the remark after the  Plethysm PH.
Furthermore, if $ Q_\lambda$ exists, its dimension is guaranteed 
to be polynomial in the length of $\lambda$ by the proof of
Theorem~\ref{tplethysm2}.

The following  is a precise form of 
Hypothesis~\ref{hph1} (b).

\begin{hypo} {\bf (PH1)} \label{hph1tilde} (GCT6)

There exists an
explicit polytope $ Q_{\lambda}$ satisfying  PH 
in Hypothesis~\ref{hphvariant}. 
Here explicit means:

\begin{enumerate}
\item  The polytope is specified by an explicit system of linear constraints,
each constraint of bitlength $\poly(\bitlength{\lambda})$
(note no dependence on $m$).

\item  The membership problem for the polytope
$ Q_{\lambda}$  also 
belongs to the complexity class $P$.
That is,  given a point $x$, whether it belongs to $ Q_{\lambda}$ 
can also be decided in $\poly(\bitlength{x},\bitlength{\lambda})$  time.
Furthermore, we assume that if $x$ does not belong to the polytope,
then the membership algorithm also gives a hyperplane separating 
$x$ from the polytope in the spirit of \cite{lovasz}. 
\end{enumerate} 
\end{hypo}

The following  addresses a relaxed form of Problem~\ref{pinvariant},
which is enough for our purposes: 

\begin{theorem} \label{tverify1}

\noindent (1)  Assuming PH1 above (Hypothesis~\ref{hph1tilde}),
 whether $V_{k \lambda}(G)$ contains a $G_{\hat g}$-invariant,
for some $k \ge 1$,
can be decided in $\poly(\bitlength{\lambda})$ time.
By Theorem~\ref{tgct2first}, this is equivalent to deciding 
whether $V_{k \lambda}(G)$ occurs in $R_V[g,m]^*$ for some $k \ge 1$. 

\noindent (2) Assuming an analogous PH1 for the subgroup 
restriction problem for $H_{\hat h} \subseteq H$,
whether $V_{k \pi}(H)$ contains an $H_{\hat h}$-invariant, for some
$k \ge 1$, can also be decided in $\poly(\bitlength{\pi},n)$ time.
By Theorem~\ref{tgct2first}, this is equivalent to deciding 
whether $V_{k \pi}(H)$ occurs in $R_W[h,n]^*$ for some $k \ge 1$. 

\end{theorem} 

\proof Similar to that of  Theorem~\ref{tlinear}. \qed 

A similar result for $R_V[f,n,m]$ or $R_{\bar W}[\bar h,n,m]$ 
would not follow from the Plethysm PH 
(or more generally, the subgroup restriction PH) 
because of the serious 
leak in Theorem\ref{tgct2second} that we discussed in 
Section~\ref{sweaktogeneral}. 
One needs a more general PH for this.
We turn to this issue next.

For any $\lambda$ of length $\le (n+1)^2$, 
let $F_{\lambda,n,m}(k)$ be the multiplicity of 
$V_{k \lambda}(\bar H)$ in $R_{\bar W}[\bar h,n,m]^*$, which 
by Theorem~\ref{tgct2second},
coincides with the multiplicity of $V_{k \lambda}(G)$
in $R_V[f,n,m]^*$. Thus $F_{\lambda,n,m}(k)$ is the same as the function 
with the same notation in Theorem~\ref{tquasi} (b). The following 
is  its restatement.

\begin{theorem} \label{tquasiaymptotic}
The function $F_{\lambda,n,m}(k)$ is an asymptotic quasi-polynomial.
\end{theorem}

The singularities of the class variety $\Delta_{\bar W}[\bar h,n,m]$ 
here are not 
normal \cite{shrawan} when $m>n$. But 
in view of the exceptional nature of the class variety
and Theorems~\ref{tplethysmqausi} and \ref{tplethysm2}, 
it may be conjectured that the deviation from
rationality and normality is small; cf. the remarks after 
Hypothesis~\ref{hph1restatement} 
below.

The following is a restatement  of Hypothesis~\ref{hph} (a).

\begin{hypo} {\bf (PH) [Positivity Hypothesis]} \label{hph2}
For every $\lambda,n,m\ge n$, 
there exists a parametrized (cf. (\ref{eqpara}))  polytope 
$P=P_{\lambda,n,m}=P_{\lambda,n,m}(k)$ 
such that 
\begin{equation} 
F_{\lambda,n,m}(k)=f_P(k)
\end{equation} 
It is also assumed that $P_{\lambda,n,m}$ has a specification as in
Hypothesis~\ref{hph} (a). 
\end{hypo}

If such $P$ exists, its  dimension is  guaranteed to be
polynomial in $n$
(by the proof of Theorem~\ref{tquasiaymptotic}) 
essentially because the dimension
of $\bar H$ is  $O(n^2)$ and does not depend on
$m$. 

The following is a restatement of Hypothesis~\ref{hph1} (a). 

\begin{hypo} {\bf (PH1)}  \label{hph1restatement}

There exists an {\em explicit} 
parametrized polytope $P_{\lambda,n,m}=P_{\lambda,n,m}(k)$ 
as in Hypothesis~\ref{hph2}.
\end{hypo} 

The meaning of explicit  here is as 
in Hypothesis~\ref{hph1}. In particular, the  polynomial bounds are
meant to be polynomial in $\bitlength{\lambda},n$ and $\bitlength{m}$,
instead of $m$. Because
$F_{\lambda,n,m}(k)$ is the multiplicity of 
$V_{k \lambda}(\bar H)$ in $R_{\bar W}[\bar h,n,m]^*$, 
$\dim(\bar H)=(n+1)^2$, which does not depend on $m$,  and $m$ occurs
only in the definition of $\bar h(\bar X)=z^{m-n} h(X)$ as a numeric parameter 
akin to the numeric parameters $\lambda_i$'s. 

Furthermore, PH1 above
implies that the deviation from quasipolynomiality of 
$F_{\lambda,n,m}(k)$ is small, specifically, 
$O(2^{O(\poly(\bitlength{\lambda},n,\bitlength{m}))})$, so that the bitlength
of the deviation is polynomial. This would mean that the deviation from
rationality and normality of the singularities of the class variety 
$\Delta_{\bar W}[\bar h,n,m]$ is also small; 
 cf.  Theorem~\ref{tquasiaymptotic} and the remarks after it.

\begin{theorem} \label{tverify2}
Assuming general  PH1  (Hypothesis~\ref{hph1restatement}), 
whether $V_{k \lambda}(G)$ occurs in $R_V[f,n,m]^*$--or
equivalently, whether $V_{k \lambda}(\bar H)$ 
occurs in  $R_{\bar W}[\bar h,n,m]^*$--for some
$k \ge 1$, can be 
decided in $\poly(\bitlength{\lambda},n,\bitlength{m})$ time.
\end{theorem}

The following is a  refined version of Theorem~\ref{tgct6}: 

\begin{theorem} (GCT6) \label{tgct6refined}
There exists a family $\{O_n=V_{\lambda_n}(G)\}$ of strong obstructions 
for the general $\#P$ vs. $NC$ problem in characteristic zero, 
for $m=2^{\log^a n}$, $a>1$ fixed, $n\rightarrow \infty$, 
assuming, 

\begin{enumerate} 
\item PH above (Hypotheses~\ref{hphvariant} and \ref{hph2}) , and
\item OH (Obstruction Hypothesis): 

For all $n\rightarrow \infty$,  there exists $\lambda$ such that 
$P_{\lambda,n,m}(k) \not = \emptyset$ for all sufficiently large $k$ and 
$Q_{\lambda}=\emptyset$.
\end{enumerate}

For an analogous result  for the $P$ vs. $NP$ problem, see GCT6.
\end{theorem}

We define $\lambda$ to be a polyhedral  obstruction (label) if it satisfies OH
here. 
In this case it is easy to see that $k \lambda$, for some $k \ge 1$,
is a strong  obstruction.

\section{Positivity and the $P$-barrier} 

Theorem~\ref{tverify1}   and \ref{tverify2} imply: 

\begin{theorem} 
Given $\lambda,n,m$, whether $\lambda$ is a polyhedral obstruction (label) 
can be decided in $\poly(\bitlength{\lambda},n,\bitlength{m})$ time
assuming PH1 (Hypotheses~\ref{hph1tilde}
 and \ref{hph1restatement}). In other words, 
the $P$-barrier for verification of obstructions
(Section~\ref{spbarrier}) can be crossed assuming
PH1. 
\end{theorem}

In conjunction with Theorem~\ref{tgct6refined}, this implies 
its stronger form:

\begin{theorem} (GCT6)
There exists an explicit (cf. Section~\ref{sflip}) 
family $\{\lambda_n\}$ of polyhedral obstructions,
for the general $\#P$ vs. $NC$ problem in characteristic zero, 
for $m=2^{\log^a n}$, $a>1$ fixed, $n\rightarrow \infty$, 
assuming PH1 and OH above. 

Analogous result holds for the $P$ vs. $NP$ problem in characteristic zero.
\end{theorem}

The strategy now is to prove PH, or rather PH1, first, and then 
prove OH using the explicit forms of the polytopes in PH1. 
GCT2,6,7,8 together give an evidence for why PH/PH1 and OH should hold. 
As far as proving OH is concerned, there is nothing that we can say at this
point since it depends on the explicit forms of the polytopes 
in PH/PH1. The remaining  question is the following.

\section{How to prove  PH1 and why should it hold?}
We now briefly describe the plan in GCT6 
to prove PH1 by generalizing the proof of LR PH1
(Theorem~\ref{tlr}) based on  the theory of standard quantum groups
 \cite{drinfeld,kashiwara,lusztig}. 

For that we need a definition.

\begin{defn} 
Let $H$ be a connected reductive subgroup of a connected reductive $G$. 
A basis $B$ of a representation $V$ of $G$ 
is called {\em positive} with respect to the $H$-action if: 

\begin{enumerate}  
\item If it $H$-compatible. This means there exists a filtration of $B$:
\[ B=B_0 \supset B_1 \supset \cdots \] 
such that $\langle B_i \rangle / \langle B_{i+1} \rangle$, where 
$\langle B_i \rangle$ denote the linear span of $B_i$, is 
isomorphic to an irreducible $H$-module. In other words, this filtration
gives a Jordan-Holder series of $V$.

\item For each standard generator $h$ of (the Lie algebra of) $H$ and
each $b \in B$, 
\[ h b = \sum_{b' \in B} c^h_{b,b'} b, \] 
where each $c^h_{b,b'}$ is a nonnegative rational. 
\end{enumerate} 
\end{defn} 

LR PH1 is  a 
consequence of the proof of a much deeper positivity result:

\begin{theorem} [LR PH0] \cite{lusztig,beilinson}
Let $H=GL_n(\C)$ embedded in $G=H\times H$ diagonally as in 
(\ref{eqdiagonallr}).
Then each irreducible 
representation of $G$ has a positive basis with respect to the $H$ action.
\end{theorem}

The proof of this result goes via the theory of the standard quantum group.
Specifically, the diagonal embedding (\ref{eqdiagonallr}) is first 
quantized \cite{drinfeld} in the form
\begin{equation}  \label{eqqdi}
H_q \rightarrow H_q \times H_q,
\end{equation}
where $H_q$ is the standard quantum group, a quantization of $H$ that
plays the same role
in quantum mechanics that the standard group $H$ plays in classical 
mechanics. (Well, (\ref{eqqdi}) is not really accurate, because what is 
quantized in \cite{drinfeld} is not $H$ but rather its universal 
enveloping algebra. We shall ignore this technicality here.) 
It is then shown that the irreducible representations of $H_q$ and 
$H_q \times H_q$ have extremely rigid {\em canonical} bases 
\cite{kashiwara,lusztig},
which are positive \cite{lusztig},
and have many other remarkable properties.
The only known proof of this
positivity 
\cite{lusztig} is based on the Riemann hypothesis over finite fields 
and the related works 
\cite{weil2,beilinson}

\begin{goal} 
Lift this LR story to the  plethysm problem (and the more general
subgroup restriction problem).
\end{goal}

In this context:

\begin{hypo} [Plethysm PH0] \label{hplph0}
Let 
\begin{equation} \label{eqpl2}
H=GL_n(\C) \rightarrow GL(V), V=V_\mu(H),
\end{equation}
be the plethysm homomorphism (\ref{eqplethysm}).
The each Weyl module $V_\lambda(G)$ has a positive basis with respect
to the $H$-action.
\end{hypo}

\begin{theorem} (GCT7)
The plethysm map (\ref{eqpl2}) can be quantized in the form 
\begin{equation} 
H_q \rightarrow G^H_q,
\end{equation}
where $H_q$ is the standard quantum group associated with 
$H$ and $G^H_q$ is a new nonstandard quantization of $G$.

A similar result holds for general connected reductive $H$ as well.
\end{theorem}

Furthermore, GCT8 gives a conjecturally correct algorithm to construct 
canonical bases of 
irreducible representations $G^H_q$ with  conjectural positivity
and other properties from which Plethysm PH0 would follow. 
Experimental evidence for 
positivity of the conjectural canonical bases in GCT8 constitutes 
the main evidence  for Plethysm PH0, and hence Plethysm PH1/PH.

\begin{figure} 
\begin{center}
\psfragscanon
\epsfig{file=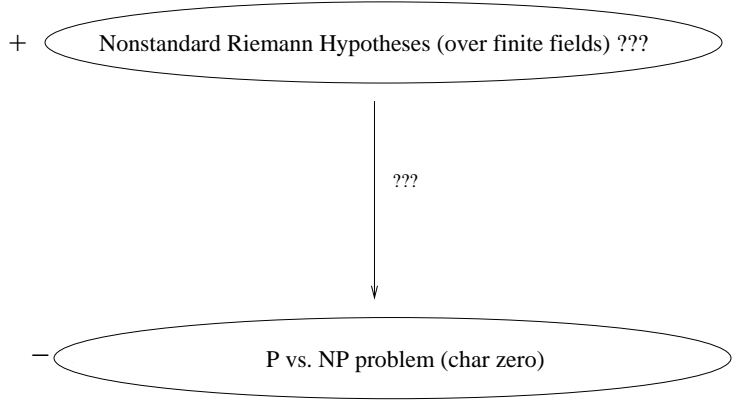, scale=.6}
\end{center}
      \caption{The ultimate goal of GCT}
      \label{fig:finalgoal}
\end{figure}

The general PH1 (Hypothesis~\ref{hph1restatement})  can be regarded as a
generalization  of the Plethysm PH1 for    the triple
(cf. Observation~\ref{obstriple})
\begin{equation} \label{eqtri}
\bar H_{\hat h} \rightarrow \bar H=SL(\bar X) \rightarrow L=GL(\bar W)
\end{equation}
associated 
with the class variety $\Delta_{\bar W}[\bar h]$,
rather than the plethysm couple (\ref{eqpl2}).
To go from the Plethysm PH1 to the general PH1, one has to similarly
quantize the triple (\ref{eqtri}) and develop an analogous theory of 
canonical bases for this quantized triple. 
But first, we have to understand the couples. Hence the Plethysm PH0/PH1
can be regarded as the heart of GCT. 
To prove the nonstandard quantum group  conjectures in GCT7,8 that arise
in this context, 
a substantial nonstandard  extension of the work 
\cite{weil2,beilinson,kazhdan,lusztig} surrounding 
the standard Riemann hypothesis over finite fields may 
be necessary; cf. Figure~\ref{fig:lift}. 
Thus the ultimate goal of GCT would be  to deduce the ultimate 
negative hypothesis of mathematics, $P \not = NP$ conjecture (in characteristic
 zero), 
from the ultimate positive hypotheses--namely, as yet unknown,
nonstandard Riemann
hypotheses (over finite fields); cf. Figure~\ref{fig:finalgoal}.

\section*{Appendix: A bit more of  representation theory} \label{sbasicdefnrepr}
Here we go into  the basic representation theory a bit more than
in Section~\ref{sbasicrepr};
in particular, we describe an explicit construction of Weyl modules. 

Let $G$ be a group.
We say that a vector space $V$ is a {\em representation}
 of $G$, or a {\em $G$-module},
 if there is 
 a homomorphism
\begin{equation} \label{eqreprmap}
\rho: G \rightarrow GL(V),
\end{equation} 
where $GL(V)$ is  the general linear 
group of invertible transformations of $V$. We denote $\rho(g)(v)$ 
by $g\cdot v$--the result of the action of $g$ on $v$. 
A $G$-subrepresentation $W \subseteq V$ is a subspace that 
is invariant under $G$; i.e., $g \cdot w \in W$ for every $w \in W$.
If $G$ is clear from the context, we just call it subrepresentation.
We say that $V$ is {\em irreducible} if it does not contain a proper 
nontrivial subrepresentation.
A {\em $G$-homomorphism}  from a $G$-module $U$ to a $G$-module $V$ is 
map $\psi: U \rightarrow V$ such that $\psi(g\cdot u)=g \cdot (\psi(u))$
for all $u \in U$.

We say that $G$ is {\em reductive} if 
every finite dimensional 
representation $V$ of $G$ is {\em completely reducible}.
This means it
can be expressed as a direct sum of irreducible 
representations in the form
\begin{equation} \label{eqcompletedecomp}
V=\bigoplus_\lambda {m_\lambda} V_\lambda(G)
\end{equation}
where $\lambda$ 
ranges over all indices (labels) of irreducible representations of $G$,
 $V_\lambda(G)$ denotes the irreducible representation of $G$ with label 
$\lambda$, and
${m_\lambda} V_\lambda(G)$ denotes a direct sum of $m_\lambda$ copies of
$V_\lambda(G)$. Here $m_\lambda$ is called the {\em multiplicity}
 of $V_\lambda(G)$
 in
$V$. It is a basic fact of representation theory that for reductive groups, 
the decomposition (\ref{eqcompletedecomp}) is 
unique; i.e., $m_\lambda$'s are uniquely defined. 
If $m_\lambda>0$, we say that $V_\lambda(G)$ {\em occurs} in $V$.

An example of a nonreductive group is a solvable group that is not abelian.
In this case a subrepresentation $W \subseteq V$ need not have a 
complement $W^\bot$ such that $V=W \oplus W^\bot$.

Every finite group is reductive. Thus
$S_n$, the symmetric group
on $n$ letters, is reductive.
A prime example of a continuous reductive group is the general linear 
group $GL_n(\C)=GL(\C^n)$, the  group of nonsingular $n\times n$ matrices,
and its subgroup the special linear group
$SL_n(\C)=SL(\C^n)$ of matrices with determinant one.
Any product of reductive groups is also reductive.  These are the only kinds 
of reductive groups that we need to know in this article. So whenever
we say reductive, the reader may wish to assume that the group is a
general or special linear group or a symmetric group or a product thereof.

We say that 
the  representation (\ref{eqreprmap}) of $GL_n(\C)$ or $SL_n(\C)$
 is polynomial if
for every $g \in G$, every entry in the matrix form  of $\rho(g)$ is
a polynomial in the entries of $g$.

Complete reducibility as in  eq.(\ref{eqcompletedecomp})
means every finite 
dimensional representation of a reductive group is composed of 
irreducible representations. These can be thought of as the building blocks
in the representation theory of reductive groups, and it is important 
to know what  these building blocks are.

For $G=GL_n(\C)$ and $SL_n(\C)$ this was done by 
Weyl  \cite{fultonrepr}.  
The polynomial irreducible representations of $GL_n(\C)$ are
in one-to-one correspondence with the  tuples
$\lambda=(\lambda_1,\ldots,\lambda_k)$ of integers, where $k \le n$ and 
$\lambda_1 \ge \lambda_2 \cdots \ge \lambda_k > 0$. 
Here $\lambda$ is called a {\em partition} of length 
 $k$ and size $|\lambda|=\sum_i \lambda_i$. Its bitlength
 $\bitlength{\lambda}$ is defined to be
the total bitlength of all $\lambda_i$'s.

Thus the polynomial irreducible representations of $GL_n(\C)$ are labelled by 
partitions $\lambda$ of length at most $n$, but any size. 
The irreducible representation corresponding to a partition 
$\lambda=(\lambda_1,\lambda_2,\ldots)$ 
is denoted by $V_\lambda(GL_n(\C))$, and  is
called a {\em Weyl module} of $GL_n(\C)$.
When $GL_n(\C)$  is clear from the context,
we shall denote it by simply $V_\lambda$.

Each partition $\lambda$ corresponds to a Young diagram, which consists of
$k$ rows of boxes, with $\lambda_i$ boxes in the $i$-th row. For example,
the Young diagram corresponding to $(4,2,1)$ is shown below: 

\[ 
\yng(4,2,1)
\]
 When  thinking of a partition,
it is helpful to think of the corresponding  Young diagram.
Thus each Weyl module is labelled by a Young diagram of height at most $n$.
This is a useful combinatorial tool for studying the Weyl modules.

A Weyl module $V_\lambda$ 
is explicitly constructed as follows.
This construction of Deyruts as well as  Weyl's original
construction are given in \cite{fultonrepr}. 
Let $Z$ be an $n\times n$ variable matrix. Let $\C[Z]$ be the ring of
polynomials in the entries of $Z$. 
It is a representation of $GL_n(\C)$. Action of a  matrix  
$\sigma \in GL_n(\C)$ 
on a polynomial  $f \in \C[Z]$ is given by
\begin{equation} 
(\sigma \cdot f)(Z)= f (Z \sigma).
\end{equation}

By a numbering (filling), we 
mean filling of the boxes of a Young diagram 
by numbers in $[n]$; for example:
\[ 
\young(1243,23,1)
\]
We call such a numbering a {\em (semistandard) tableau} if the numbers
are strictly increasing in each column and weakly increasing in all rows;
e.g.
\[ 
\young(1233,23,4)
\]

The partition corresponding to the Young diagram of a numbering
is called the {\em shape} of the numbering.

With every  numbering $T$, we associate a polynomial $e_T \in \C[Z]$,
which is a product of minors for each column of $T$. The $l\times l$ 
minor $e_c$ for a column $c$ of length $l$
is formed by the first $l$ rows of $Z$ and the columns indexed 
by the entries $c_j$, $1 \le j \le l$, of $c$. Thus $e_T=\prod_c e_c$,
where $c$ ranges over all columns in $T$. The Weyl module 
$V_\lambda$ is the subrepresentation of $\C[Z]$  spanned by $e_T$,
where $T$ ranges over all 
numberings of shape $\lambda$ over $[n]$. 
Its one possible basis is given by $\{e_T\}$, where $T$ ranges over 
semistandard tableau of shape 
$\lambda$ over $[n]$.

Let $B \subseteq GL_n(\C)$
be the subgroup of upper triangular matrices. It is called 
the {\em Borel subgroup} of $GL_n(\C)$.
An element  $v_\lambda \in V_\lambda$ is called a
{\em highest weight vector} if it
is an eigenvector for the action of each $b \in B$.
It is easy to show that $V_\lambda$ has a unique highest weight vector,
upto a constant multiple:  it is $e_{T_0}$, where
$T_0$ is the canonical tableau whose $i$-th row contains only $i$'s, for 
each $i$; e.g.
\[ 
\young(1111,22,3)
\]

Let $P \subseteq GL_n(\C)$ be the subgroup of upper block triangular matrices,
where the sizes of the blocks are fixed. For example: 

\[ 
\left[\begin{array} {llllll}
* & * & * & * & * & * \\
* & * & * & * & * & * \\
0 & 0 & * & * & * & * \\
0 & 0 & * & * & * & * \\
0 & 0 & 0 & 0 & * & *  \\
0 & 0 & 0 & 0 & * & *
\end{array}
\right]
\]

Such subgroups are called {\em parabolic}. 
Let $P_\lambda$ be the (projective) stabilizer of the highest weight 
vector $v_\lambda=e_{T_0}$; i.e.,
the set of all $\sigma \in GL_n(\C)$
 such that $\sigma \cdot v_\lambda = c(\sigma)
v_\sigma$, for some complex number $c(\sigma)$.
Then it is easy to show that $P_\lambda$ is parabolic, where the 
sizes of the blocks are completely determined by $\lambda$.

The irreducible representation of $GL_n(\C)$ corresponding to 
the Young diagram that consists of just one column of length $n$ is 
the determinant representation: $g \rightarrow \det(g)$. When restricted 
to the subgroup $SL_n(\C) \subseteq GL_n(\C)$ this becomes trivial. 
More generally, $V_\lambda(GL_n)$ and $V_{\lambda'}(GL_n)$ give the
same representation of $SL_n(\C)$ if $\lambda'$ is obtained from
$\lambda$ by removing columns of length $n$. Hence,
irreducible polynomial representations of $SL_n(\C)$  are in one to
one correspondence with partitions of length less than $n$, and 
are obtained from the ones of $GL_n(\C)$ by restriction.

<


\begin{thebibliography}{[Welzl]}

\bibitem{aaronsonformal} S. Aaronson, Is P versus NP formally independent? 
Bulletin of the EATCS 81: 109-136 (2003). 

\bibitem{bharat} B. Adsul, private communication.

\ignore{\bibitem{agarwal} M. Agrawal et al, Reducing the complexity of reductions,
preprint, 2001.



\bibitem{aaronson} S. Aaronson, A. Wigderson, Algebrization: a new barrier in 
complexity theory, ACM transactions on computing theory, 1(1), 2009.}

\bibitem{boris} B. Alexeev, J. Tsimerman, A direct proof of Mulmuley's
weak $\#P$ versus $NC$ result, manuscript.


\bibitem{ariki} S. Ariki, J. Matsuzawa, I. Terada, Representations of Weyl
groups on zero weight spaces of ${\cal g}$-modules, Algebraic and topological 
theories, pp. 546-568 (1985). 

\ignore{\bibitem{solovay} T. Baker, J. Gill, R. Soloway, Relativization of the $P=?NP$ question, 
SIAM J. Comput. 4, 431-442, 1975.}



\bibitem{beilinson} A. Beilinson, J. Bernstein, P. Deligne, Faisceaux pervers, Ast\'erisque 
100, (1982), Soc. Math. France.

\ignore{\bibitem{benor} M. Ben-Or, Lower bounds for algebraic computation trees,
proceedings of the STOC, 83, 80-86.

\bibitem{sipser} R. Boppana, M. Sipser: The complexity of finite functions,
Handbook of Theoretical Computer Science, vol. A,
Edited by
J. van Leeuwen, 
North Holland, Amsterdam, 1990, 757--804. 
}


\bibitem{boutot} J. Boutot, Singularit'es rationelles et quotients par les 
groupes r'eductifs,
Invent. Math.88, (1987), 65-68.

\bibitem{brion} M. Brion, On the general faces of the moment polytope, 
IMRN International Mathematics Research Notices, No. 4, 1999.

\bibitem{cook}  S. Cook: The complexity of theorem-proving procedures. 
Proceedings of the third annual ACM Symposium on Theory of Computing.
151-158. (1971). 


\bibitem{dehy} R. Dehy, Combinatorial results on Demazure modules, J. of Algebra 205, 505-524 
(1998). 

\bibitem{weil2} P. Deligne, La conjecture de Weil II, Publ. Math. Inst. Haut. \'Etud. Sci. 52,
(1980) 137-252. 

\bibitem{delignet} P. Deligne, Categories tannakiennes, in 
The Grothendieck Festschrift, Volume 2, 11-195, Birkhauser, 19990. 

\bibitem{drinfeld}
V. Drinfeld, Quantum groups, Proc. Int. Congr. Math. Berkeley, 1986,
vol. 1, Amer. Math. Soc. 1988, 798-820.

\bibitem{flenner} H. Flenner, 
Rationale quasi-homogene singularit\"aten, Arch. Math. 36
(1981), 35-44.

\bibitem{fultonrepr} W. Fulton, J. Harris, Representation theory,
 A first course, Springer, 1991.




\bibitem{lovasz} M. Gr\"otschel, L. Lov\'asz, A. Schrijver, 
Geometric algorithms and combinatorial optimzation, Springer-Verlag,
1993.


\bibitem{hironaka} H. Hironaka, Resolution of singularities of an
algebraic variety over a field of characteristic zero, Ann. of 
Math (2), 79: 109-273.


\bibitem{karp} R. Karp: Reducibility among combinatorial problems. 
R. E. Miller and J. W. Thatcher (eds.) Complexity of computer computations,
Plenum Press, New York, 1972, 85-103. 


\bibitem{kashiwara}  M. Kashiwara, On crystal bases of the $q$-analogue of
 universal enveloping algebras, Duke Math. J. 63 (1991), 465-516.


\bibitem{kazhdan} D. Kazhdan, G. Lusztig, Schubert varieties and Poincare
duality, Proc. Symp. Pure Math., AMS, 36 (1980), 185-203. 

\bibitem{kempf} G. Kempf: Instability in invariant theory, Annals of 
Mathematics, 108 (1978), 299-316. 



\bibitem{kirillov} A. Kirillov, An invitation to the generalized saturation
conjecture, math. CO/0404353., 20 Apr. 2004.


\bibitem{knutson} A. Knutson, T. Tao, The Honeycomb model of $GL_n(\C)$
tensor products I: proof of the saturation conjecture, J. Amer. Math. Soc, 12,
1999, pp. 1055-1090.

\bibitem{levin} A. Levin: Universal sequential search problems. 
Problems of information transmission  (translated from 
Problemy Peredachi Informatsii (Russian)) 9 (1973). 


\bibitem{lusztig} G. Lusztig, Canonical bases arising from 
quantized enveloping algebras, J. Amer. Math. Soc. 3, (1990), 447-498.


\bibitem{mayr} E. Mayr, and A. Meyer, The complexity of the word
problems for commutative semigroups and polynomial ideals, 
Advances in mathematics, 46 (3): 305-329, 1982.


\bibitem{ressayre} T. Mignon, N. Ressayre, A quadratic bound for the 
determinant and permanent problem, International Mathematics 
Research Notices (2004) 2004: 4241-4253.

\bibitem{GCTpram} K. Mulmuley: 
Lower bounds in a parallel model without
bit operations, The SIAM Journal On Computing, vol. 28, no. 4, 1999. 



\bibitem{GCTflip} K. Mulmuley, On P vs. NP, geometric complexity theory,
and the flip I, Technical report TR-2007-16,
computer science department, The university of Chicago, September 2007;
revised version under preparation.
This and the following GCT papers are available at http://ramakrishnadas.cs.uchicago.edu. 


\bibitem{GCTexplicit} K. Mulmuley, On P vs. NP, Geometric Complexity Theory,
Explicit proofs, and the Complexity Barrier, under preparation, to be
available at the above website soon. 




\bibitem{GCT1} K. Mulmuley, M. Sohoni, Geometric complexity theory I:
an approach to the $P$ vs. $NP$ and related problems, 
SIAM J. Comput., vol 31, no 2, pp 496-526, 2001.

\bibitem{GCT2} K. Mulmuley, M. Sohoni, Geometric complexity theory II: 
towards explicit obstructions for embeddings among class varieties, 
SIAM J. Comput., Vol. 38, Issue 3, June 2008.

\bibitem{GCT2append} K. Mulmuley, 
An appendix to Geometric Complexity Theory II,  technical report, 
computer science department, the university of chicago, February, 2009.




\ignore{
\bibitem{GCT3} K. Mulmuley, M. Sohoni, Geometric complexity theory III,
on deciding positivity of Littlewood-Richardson coefficients, cs. ArXiv 
preprint cs. CC/0501076 v1 26 Jan 2005.


\bibitem{GCT4} K. Mulmuley, M. Sohoni, Geometric complexity theory IV: 
quantum group for the Kronecker problem, cs. ArXiv preprint cs. CC/0703110,
March, 2007.

\bibitem{GCT5} K. Mulmuley, H. Narayanan, Geometric complexity theory V:
on deciding nonvanishing of a generalized Littlewood-Richardson coefficient,
Technical Report TR-2007-05,
computer science
department, The University of Chicago, May, 2007.
Available at: http://ramakrishnadas.cs.uchicago.edu}

\bibitem{GCT6} K. Mulmuley, Geometric complexity theory VI: The flip,
Revised version under preparation, 
Earlier version: Technical Report TR-2007-04, computer science
department, The University of Chicago, September, 2007.




\bibitem{GCT7} K. Mulmuley, Geometric complexity theory VII:
Nonstandard quantum group for the plethysm problem,
Technical Report TR-2007-14, computer science
department, The University of Chicago, September, 2007.




\bibitem{GCT8} K. Mulmuley, Geometric complexity theory VIII: 
On canonical bases for the nonstandard 
quantum groups,  Technical Report TR 2007-15, computer science department,
The university of Chicago, September 2007. 


\bibitem{GCT11} K. Mulmuley, Geometric complexity theory XI: 
on the flip over  fields of positive characteristic, under preparation.

\bibitem{shrawan} Shrawan Kumar, private communication.

\ignore{\bibitem{sarnak} A. Lubotzky, R. Phillips, P. Sarnak: Ramanujan graphs,
Combinatorica 8 (1988), 261-277. 
}



\bibitem{mumford}  D. Mumford, J. Fogarty, F. Kirwan: 
Geometric invariant theory. Springer-Verlag, 1994. 

\bibitem{popov} V. Popov, E. Vinberg, Invariant theory, 
in Encyclopaedia of Mathematical Sciences,
Algebraic Geometry IV, Eds. A. Parshin, I. Shafarevich, 
Springer-Verlag, 1989.

\ignore{\bibitem{monotone} A. Razborov, Lower bounds on the monotone complexity of 
some boolean functions, Doklady Akademii Nauk SSSR 281 (1985), 798-801.}


\bibitem{rudich} A. Razborov, S. Rudich, Natural proofs, J.
 Comput. System Sci., 55 (1997), pp. 24-35. 



\bibitem{smith} K. Smith, F-rational rings have rational 
singularities, Amer. J. Math. 119 (1997).


\bibitem{stanleyenu} R. Stanley, Enumerative combinatorics, vol. 1, 
Wadsworth and Brooks/Cole, Advanced Books and Software, 1986.



\bibitem{valiant} L. Valiant: 
The complexity of computing the permanent. Theoretical 
Computer Science 8 , 189-201 (1979).



\end{thebibliography}
\end{document}